\documentclass[fleqn,usenatbib,onecolumn]{mnras}

\usepackage{mathptmx}
\usepackage[T1]{fontenc}
\usepackage{ae,aecompl}

\usepackage{graphicx}	% Including figure files
\usepackage{amsmath}	% Advanced maths commands
\usepackage{amssymb}	% Extra maths symbols
\usepackage{color}
\title{Blob formation and ejection from the radiative inefficient accretion 
flow around massive black hole}

\author[T.L. Zhao et al.]{
Tian-Le Zhao,$^{1,2}$\thanks{E-mail:tianle@mail.ustc.edu.cn}
Ye-Fei Yuan,$^{1,2}$\thanks{yfyuan@ustc.edu.cn}
and Rajiv Kumar$^{1}$
\\
$^{1}$CAS Key Laboratory for Research in Galaxies and Cosmology, Department of Astronomy, University of Science and Technology of China, Hefei 230026, China \\
$^{2}$School of Astronomy and Space Sciences, University of Science and Technology of China, Hefei 230026, China\\
}

\date{Accepted XXX. Received YYY; in original form ZZZ}

\pubyear{2020}
\begin{document}
\label{firstpage}
\pagerange{\pageref{firstpage}--\pageref{lastpage}}
\maketitle

\begin{abstract}
We study the small scale magnetic reconnection above the radiative inefficient accretion flow around massive black hole via 2D magnetohydrodynamics (MHD) numerical simulation, in order to model the blob formation and ejection from the accretion flow around Sgr A*.
	The connection of both the newly emerging magnetic field and the pre-existing magnetic field is investigated to check whether blobs could be driven in the environment of black hole accretion disc. After the magnetic connection, both the velocity and temperature of the plasma can be comparable to the inferred physical properties at the base of the observed blob ejection.
For illustration, three small boxes which are located within 40 Schwarzschild radii from the central black hole are chosen as our simulation areas. At the beginning of the reconnections, the fluid is pulled toward the central black hole due to the gravitational attraction and the current sheet produced by the reconnection is also pulled toward the same direction, consequently, the resulting outflows move both upwards and towards the symmetry axis of the central black hole. Eventually, huge blobs appear, which supports the catastrophe model of episodic jets \citep{2009MNRAS.395.2183Y}. 
It is also found that the closer to the black hole the magnetic connection happens, the higher the converting efficiency of the magnetic energy into the heat and kinetic energy. For these inner blobs, they have vortex structure due to the Kelvin–Helmholtz instability, which happens along the current sheet separating the fluids with different speed. 
	
\end{abstract}


%%%%%%%%%%%%%%%%%%%%%%%%%%%%%%%%%%%%%%%%%%%%%%%%%%

%%%%%%%%%%%%%%%%% BODY OF PAPER %%%%%%%%%%%%%%%%%%

\section{Introduction} % (fold)
\label{sec:introduction}
Jets are ubiquitous in the center of the accreting objects, such as, active galactic nuclei (AGN), black hole $X-$ray binaries, gamma-ray burst and young stellar objects. They can be observed
at multi-wavelength bands, from radio to gamma-ray. Some observations indicate that 
AGN jets can transport energy and gas from their central regions (sub-parsec scale) to 
their host galaxy (Mpc scale), and feed back the evolution of their host galaxy. 
Owing to the  limitation of angular resolution of present 
ground-based and space telescopes \citep{2016ApJ...817...96G}, the base of jet has not been 
directly observed \citep{2016Galax...4...54F}.  Event Horizon Telescope (EHT), in principle, could resolve the base of jets of Sgr A*, M87*, or even nearby AGNs in the near future \citep{3C279-EHT}.
So the numerical simulation is an important method to study the production of AGN jets. 
In the past 40 yr, there have been significant progresses in the numerical magnetohydrodynamics (MHD) 
simulation on jet formation, both in the numerical techniques  and the detailed theoretical models. 
The feasibility of numerical simulation is confirmed by Norman and his collaborators 
who performed the simulation of the jet formation successfully in the early 1980s \citep{1982A&A...113..285N}. 
%Therefore, the theoretical model of AGN jet has become the main direction of numerical simulation in the past 40 years. 
The bottom line here is that the jet is produced at the scales of a few gravitational radii ($R_\mathrm{s}$) of 
the central black hole (BH) and extends to hundreds of kiloparsecs. 
%($R_{BH}\approx10^{-4}(M_{BH}/10^9M\odot)pc$, here $M_{BH}$ is the mass of the central black hole).  

In the AGNs, the central objects are rotating supermassive BHs (SMBHs), which are supplied by the gas from the dust torus. It is generally believed that the relativistic jets are produced in the inner region of the accretion disc of SMBHs, and the formation, collimation of jets, and their acceleration up to relativistic velocity can be explained by the MHD processes. The mechanism for jet production is still under debate. In the famous Blandford–Payne (BP) model, the large-scale magnetic field attached to the accretion disc accelerates a part of the gas near the disc surface to form an outflow through magnetic centrifugation  \citep{1982MNRAS.199..883B, 2003ApJ...596.1080V,2004ApJ...605..656V,2006MNRAS.367..375B,2009ApJ...698.1570L}.Another famous mechanism is so-called Blandford–Znajek (BZ) process, in which the jet energy is extracted from the internal energy of a rotating BH \citep{1977MNRAS.179..433B,1992ApJ...386..455H}. However, there is another way to drive the jet, the magnetic reconnection mechanism which is proposed in  \citet{2009MNRAS.395.2183Y}. 
In their model,  the catastrophe model proposed by \citet{2000JGR...105.2375L} is generalized to 
explain the intermittent jet production in the BH accretion disc \citep{2009MNRAS.395.2183Y}. 
However, the Lin-Forbes model is not only a catastrophe model for jet ejection, but also 
the models for standard-jet and the blow-out jet \citep{2010ApJ...720..757M}. The standard-jet is produced by the magnetic reconnection between the newly emerged magnetic flux and the pre-existing magnetic field. It has a simple magnetic field structure with only one reconnection X-type point, which follows the ‘standard image’ of the jets, so it is called standard-jet. In contrast, the blow-out jet has a complex magnetic field structure with multiple reconnection X-points.

It is well known that the solar atmosphere/corona and the inner region of the accretion discs of AGNs are in plasma states, magnetic activities, such as magnetic reconnection, are the dominant physical processes, so the jet formation and acceleration due to the magnetic connection might be very possible. The emerging of magnetic flux generally results in the electric current. And in the process of magnetic reconnection, there is a dissipative region called current sheet where magnetic energy is converted into heat energy and kinetic energy of plasma \citep{2019arXiv190404777K}, 
which also leads to a lot of instabilities during this process \citep{2012MNRAS.422.1436O,2016ApJ...824...48S}. 
 Previous studies have shown that the fast magnetic reconnection play an important role in the accretion disc of BH \citep{2019arXiv190404777K}, for instance, the turbulent and fast reconnection 
of accretion disks can explain the $\gamma$-ray from the center of the BHs \citep{1999ApJ...511..193V,1999ApJ...517..700L,2015ApJ...802..113K,2015ApJ...799L..20S}. In addition of the catastrophe model, 
there is another model called jets in jet.  In a main and large jet, blobs are generated due to the 
magnetic reconnection, these blobs are considered as the observed mini-jets 
\citep{2009MNRAS.395L..29G}. 
%In addition, some authors have proposed that energy dissipation by magnetic reconnection in an alternating magnetic field is another energy conversion mechanism \citep{2009MNRAS.395L..29G,2010ApJ...725L.234L}. On the other hand, the magnetic reconnection mechanism plays an important role in the accretion process during the jet formation \citep{2014MNRAS.440.2185D,2018MNRAS.478.5404S,2018ApJ...859...28Q}.

The reconnection is related to the instabilities. If it is energetically favorable, then 
it tends to occur naturally and  also can generate own flow \citep{2016ASSL..427.....G}. 
Two bent magnetic field lines can be formed when the field lines break and reconnect. 
%{They want to straighten and release their energy but this process need the energy released by the reconnected field lines must overcome the energy it takes to bend the magnetic fields that have not yet reconnected.} 
They have tendency to straighten but this process needs to release their energy by reconnecting field lines, which overcome the energy of it that have taken to bend the field lines before the reconnection. If the driving force to straighten the magnetic field is large enough, then this process will continue even in the absence of force. This is known as the tearing mode instability(TMI) mechanism \citep{1963PhFl....6..459F}, which is the linear phase of magnetic reconnection. In addition, magnetic reconnection can also cause other instabilities such as current-driven instability (CDI), Kelvin-Helmholtz instability (KHI). The KHI was found on a shear plane with relative motions between two fluids by Kelvin (1871) and Helmholtz (1868) one and a half centuries ago, it looks like a vertex structure, 
and it results from the flow shear in a fluid or unmagnetized plasma, and it can be stoped by the magnetic fields parallel or anti-parallel\citep{2016ASSL..427.....G}.   In the magnetically dominated area such as strong toroidal magnetic field,
the turbulent reconnection can generate currents, which causes current-driven instability (CDI), this instability develops if the length of the plasma column is long enough for the field lines to go around the cylinder at least once\citep{2011IAUS..275...41H}. These instabilities can result in the observational imprints such as radio knots, transverse structure, bright spots, 
blobs, and so on.
%such as the recent detection of blobs near the last stable circular orbit of the massive black hole SgrA*\citep{2018A&A...618L..10G}.
So these characteristics can be used to constrain the parameters of jets \citep{2011IAUS..275...41H} and can 
help to study the formation and evolution of the jets. This analysis can probe the physical conditions in the jets of several sources,
such as M87\citep{1989ApJ...340..698O,2003NewAR..47..629L,2011ApJ...735...61H}, S50836+710\citep{1998A&A...340L..60L,2007A&A...469L..23P,2012ApJ...749...55P}, 3C273 (Lobanov \& Zensus 2001), 3C120\citep{2001ApJ...556..756W,2003NewAR..47..645W,2005ApJ...620..646H} and BL Lac\citep{2015ApJ...803....3C}. There are lots of studies focus on the stability of jet flow in the pure polar magnetic field \citep{2007ApJ...662..835M, 2009ApJ...692L..40Z,2011ApJS..193....6B}.
Purpose of this work is to expound the basic properties of the turbulence of MHD driven by KHI (magnetic field dynamics and energy spectrum). \newline

Recently, {\it Very Large Telescope Interferometer} (VLTI) observed several near-infrared superflares which might be from the hot spots at about 10 to 15 Schwarzschild radii from Sgr~A* \citep{2018A&A...618L..10G, 2020ApJ...891L..36G}, it is believed that these hot spots are likely due to the relativistic magnetic reconnection event. 
The details of the jets have been observed (e.g. \citep {2016A&A...595A..54M, 2019arXiv190404777K}), which
can be also seen in the simulations of the small-scale jet. 
So numerical simulation of the process of magnetic reconnection and the relevant instabilities  \citep{2019arXiv190404777K} 
can be used to understand these observed physical phenomena \citep{2020arXiv200603657D,2020arXiv200514251B,2020arXiv200603658P}.
There are many numerical simulations on the details of jets. At first, the numerical simulation of the 
jet produced by the accretion of BH with rapid rotation shows that it has obvious instability\citep{2006MNRAS.368.1561M,2009MNRAS.394L.126M}. 
This phenomenon has stimulated a series of numerical studies on jet instability. \citet{2007ApJ...662..835M} focused on the magnetized  rotation instability of the spine-sheath jets with purely poloidal magnetic fields. In other simulations, we notice that during the process of jet formation, 
the shear motion of jet and the rotation of the plasma also produce a series of instabilities\citep{2009ApJ...700..684M,  2011ApJ...734...19M, 2012ApJ...757...16M}. The CDI and KHI can also occur in the process of the magnetic reconnection\citep{2014ApJ...784..167M,2016ApJ...824...48S}. 
In \citet{2018ApJ...860..121F}, the synchrotron radiation is calculated based on a  series of model\citep{2016ApJ...831..163M}. 
At present, most of the numerical simulation work on flares is to check the production of a bright spot on an accretion disc. Some numerical simulations of magnetic reconnection are also devoted to the study of the nature of magnetic reconnection or the instabilities, and most of the simulations are under the spiral field and cylindrical coordinates.  Our MHD numerical simulation work is to study the process of magnetic reconnection in a small scale area above the disc, as in the standard model  \citep{2010ApJ...720..757M}. We have introduced the anisotropic heat conduction and the results show that the instability can cause the blob formation, which is different from the previous works which focus on the conditions affecting the instabilities\citep{2009ApJ...700..684M,  2011ApJ...734...19M, 2012ApJ...757...16M}.
% In this work, we purpose to connect the properties of MHD jet and the structure of the extragalactic jet observed at the parsec scale.\newline

In our previous work, we successfully simulated the production of standard-jet in the low solar corona \citep{2010ApJ...720..757M}. Our main result is the formation of bright blobs in the
corona due to the K-H instability and tears-mode instability \citep{2017ApJ...841...27N,2018RAA....18...45Z}. Based on our previous work, we investigate the blob formation and 
ejection from the accretion disk of supermassive black hole in this work.
Section 2 gives a very brief introduction of the numerical model,
the numerical results are shown in \S 3, and the main conclusions are summarized 
in  \S 4.

% section introduction (end)

\section{Numerical Model} % (fold)
\label{sec:Numerical Model}
\subsection{Accretion disk model}
\label{disk}
Sgr~A*, the massive black hole resides at the center of our galaxy, is a low-luminous AGN, and it accretes the surrounding gas with a low accretion rate due to the insufficient gas supply 
\citep[and references therein]{2014ARA&A..52..529Y}. 
Multi-wavelength observations of Sgr~A* from radio to X-ray is very successfully
modeled by the advection-dominated accretion flow (ADAF) model
\citep{1995Natur.374..623N, 1998ApJ...492..554N, 2003ApJ...598..301Y, 2004ApJ...606..894Y}.
There might exist jet in Sgr~A*. 
\citet{2002A&A...383..854Y} analyzed the Chandra observation of the radio source of the Sgr~A* center and proposed a jet on the accretion disk model to explain the observation for Sgr~A*. The Sgr~A* has soft spectrum and rapidly variable source, so they found that the features of X-ray spectrum can be explained by ADAF. 
They also found that the  synchrotron emission from the jet dominates over the bremsstrahlung from the ADAF, and the timescale is very short, so the predicted X-ray slope and the radio spectrum is fitted with the observations.
%The accretion mode is so-called Advection Dominated Accretion Flow (ADAF).
So we have chosen the atmosphere of the ADAFs as an initial boundary conditions in our simulation.
The Sgr~A* often shows bright, episodic $X$-ray and near-infrared flares observationally \citep{2018A&A...618L..10G}. 
 Some researches showed the flares could be consistent with hotspots 
 in the inner orbit of the AGN \citep{2020arXiv200514251B, 2020arXiv200603657D, 2020arXiv200603658P}.  The origin of these flares is not well understood. So here we would like to investigate the origin of these flares 
with the MHD simulation. The mass of the BH is taken to be 
$M_\bullet=4\times10^6 M_\odot$ and the corresponding Schwarzschild radius is 
%$R_\mathrm{s} = 1.18\times10^{10}$~m ($R_\mathrm{s}\approx3\times10^{3}(M_\bullet/M\odot)$~m). 
$R_\mathrm{s}  \approx3\times10^{3}(M_\bullet/M\odot) $~m. The global structure of ADAF of Sgr~A* 
could be found in \citet{apj699_722} and references therein.
The effect of electrons is not considered in our model because comparing with the effects of proton,  
the influence of electrons on the magnetic reconnection  is so small that it can be ignored, 
and thus only the temperature of ion is considered. 

\subsection{MHD equations}
The single-fluid MHD equations including gravity and thermal conduction is used to set up the model:
\begin{eqnarray}
\partial_t \rho &=& -\nabla \cdot \left(\rho \mathbf{v}\right),                                                                \\ 
\partial_t \mathbf{B} &=& \nabla \times \left(\mathbf{v} \times \mathbf{B}-\eta\nabla \times \mathbf{B}\right), \label{e:induction}\\
\partial_t (\rho \mathbf{v}) &=& -\nabla \cdot \left[\rho \mathbf{v}\mathbf{v}
+\left(p+\frac {1}{2\mu_0} \vert \mathbf{B} \vert^2\right)\mbox{\bfseries\sffamily I} \right]  \nonumber \\
& &+\nabla \cdot \left[\frac{1}{\mu_0} \mathbf{B} \mathbf{B} \right] + \rho \mathbf{g}, \\
\partial_t e &=& - \nabla \cdot \left[ \left(e+p+\frac {1}{2\mu_0 }\vert \mathbf{B} \vert^2\right)\mathbf{v} \right] \nonumber\\
& &+\nabla \cdot \left[\frac {1}{\mu_0} \left(\mathbf{v} \cdot \mathbf{B}\right)\mathbf{B}\right] + \nabla \cdot \left[ \frac{\eta}{\mu_0} \textbf {B} \times \left(\nabla \times \mathbf{B}\right) \right] \nonumber\\
& & -\nabla \cdot \mathbf{F}_\mathrm{C}+\rho \mathbf{g} \cdot \mathbf{v},   \\
e &=& \frac{p}{\Gamma_0-1}+\frac{1}{2}\rho \vert  \mathbf{v} \vert^2+\frac{1}{2\mu_0}\vert \mathbf{B} \vert^2,          \\
%p &=& \frac{2\rho k_{\mathrm{B}}T}{m_\mathrm{i}} .
p &=& \frac{2\rho k_B T}{m_i}.
\end{eqnarray}
% section section_name (end)
Here, the magnetic field, plasma density, total energy density, velocity and the gas pressure is represented by $\mathbf{B}$, $ \rho $, $ e $, $ v $ and $ p $, respectively. $ m_i=1.67\times10^{-27}kg $ is the mass of proton and $k_B=1.38\times10^{-23}JK^{-1}$ is the Boltzmann constant, 
$\mathbf{g}=g_x~\mathbf{e}_x$ is the acceleration due to the gravity along the radial direction. For simplicity, We consider the pseudo-Newtonian potential $\Phi= - \frac{GM}{r-rs}$ around the Schwarzschild BH
to mimic the effects of general relativity. The anisotropic heat conduction flux $F_C$ (e.g., see also Spitzer 1962)  and the temperature-dependent magnetic diffusivity $ \eta $ are set the same as those in our previous work\citep{2017ApJ...841...27N,2018RAA....18...45Z}.
The anisotropic thermal conduction leads  to the 
 transfer of the heat from the reconnection X-point into the plasmoids, 
 and this process can increase the rate of reconnection and the efficiency of the conversion of magnetic energy into the thermal energy and kinetic energy of the bulk motions\citep{2012ApJ...758...20N}. Therefore, the heat conduction $F_C$ is important in the direction parallel to the magnetic field.\newline

The NIRVANA3.8 code\footnote{See: http://nirvana-code.aip.de} is used to study the magnetic reconnection and other MHD phenomenon. This code has been nicely described in the previous works, such as, \citet{2017ApJ...841...27N,2018RAA....18...45Z}). The adaptive mesh refinement method in this simulation is the same as  in the paper by \cite{2017ApJ...841...27N}, the base-level grid is $40 \times 40$ and we use the AMR levels is 9, that is the highest refinement level grid  is $(40\times2^9) \times (40\times2^9)$.\newline

\subsection{Initial Condition}  
\label{sec:Initial Condition}
The parameters of the accretion flow at a particular radius are set the same as those in the corresponding 
radius of stationary ADAF in this simulation, which is described in the \S \ref{disk}. For the comparative study, 
we select three different radii ($R_\mathrm{in}=10R_\mathrm{s}$, $R_\mathrm{in}=30R_\mathrm{s}$ and $R_\mathrm{in}=40R_\mathrm{s}$) at the accretion disk,
which are listed with the corresponding parameters of accretion flow in Table~\ref{table1}.  
The position of the Inner-most stable circular orbit (ISCO) is labelled as a reference point 
in each figure, that is $R_\mathrm{ISCO}=3R_\mathrm{s}$. 
And the size of each simulation box is set to be $300L_0 \times 300L_0 $ (here $ L_0=10^6 $~m),
which is roughly  $R_\mathrm{s}/30\times R_\mathrm{s}/30$.  And the simulation boxes are as high as $H_0$
above the disk surface. The strength of the initial 
magnetic field is taken to be  $B_{x0} = - 0.6b_0$ and $ B_{y0} = - 0.8b_0$, where $b_0=16G$.  \newline

\begin{figure}
	\centering
	\includegraphics[width=0.80\textwidth, angle=0]{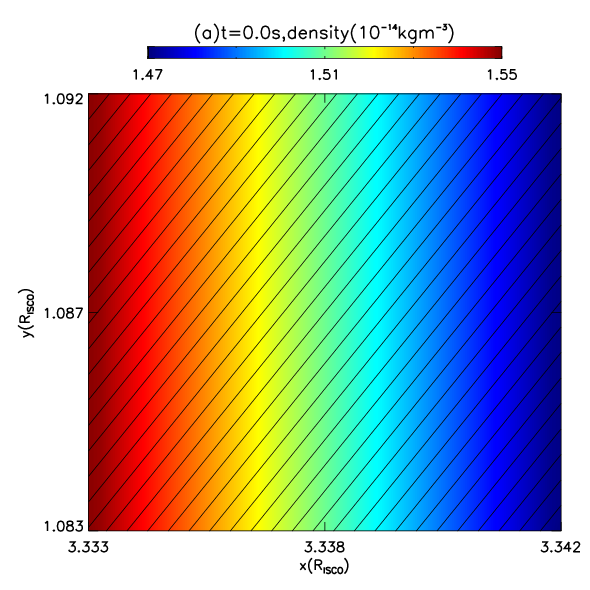}
	\caption{ The initial distribution of density and configuration of the magnetic field distribution in Case I. The solid black lines represent the magnetic fields.}
	
	\label{ini}
\end{figure}

The simulation area is tilt over the accretion disk equatorial plane, and it set as left $(x = R_\mathrm{in})$, right $(x =R_\mathrm{in}+300L_0)$, lower $(y =H_0)$ and upper $(y =H_0+ 300L_0)$ points of the box, here $H_0$ is the
height of the accretion flow.  The plasma velocity is $0$ at the time $t=0$.
Here $x$ is the radial distance from the black hole center, $R_\mathrm{in}$ is the inner radius of the simulation box. We assume that acceleration due to gravity is constant and it is approximated as a constant vector $\mathbf{g}=g_x~\mathbf{e}_x$ along the X-direction in the box because the simulation area is very small relative to the disk size. The distribution of gas density $\rho$, the pressure ($P_0$) and the thermal energy density ($E_{t0}$) on left boundary of the simulation are follows:
\begin{equation}
\rho=\rho_{0}\times \exp\left(-\frac{{(x*\sin\theta)}^2}{{H_0}^2}\right)
\label{rho},
\end{equation}
here, $\sin\theta=H_0/R_\mathrm{in}$, and
\begin{equation}
P_0=\frac{2\rho_0 k_B T_0}{m_i} , 
\end{equation}

\begin{equation}
E_{t0}=\frac{P_0}{\gamma-1}, 
\end{equation}
here $T_0$ is  the temperature of the Leftmost in simulation and the $\gamma=5/3$ is the adiabatic index, so the distribution of the thermal energy will be derived from hydrostatic equilibrium formula as:
\begin{equation}
\rho\nabla\Phi=-\nabla P.
\end{equation}
The initial mass density and the magnetic field are shown in Fig.~\ref{ini}. The initial plasma velocity is set to be zero and there is no buoyancy force. 
The average plasma-$\beta$ and the maximum parameter $\sigma=B^2/4\pi\rho c^2$\citep{2020arXiv200514251B} in three cases are shown in Table~\ref{table1}.

\subsection{Boundary Condition}  \label{sec:Boundary Condition}
Two extra layers with the ghost grid cell are set at each boundary, and the boundary conditions are the same as those in the previous work \citep{2017ApJ...841...27N,2018RAA....18...45Z}. The outflow boundary condition is set on left $ (x = R_\mathrm{in}) $, right $ (x =  R_\mathrm{in}+300L_0) $ and upper $  (y =H_0+ 300L_0) $ boundaries. The condition of being divergence-free for the magnetic field requires continuity in the normal component of the magnetic field on the boundary, which can be used to extrapolate the normal component through the boundary. Two ghost layers below the physical is inserted in bottom boundary $ y = H_0 $. The gradient of the plasma velocity is vanished at the bottom boundary. The magnetic field inside the two layers with the ghost grid cells is set as:
\begin{equation}
b_{xb}=-0.6b_0+\frac{100L_0(y-y_0)b_1 f}{[(x-x_0)^2+(y-y_0)^2]}
\left[\mathrm{tanh}\left(\frac{x-170L_0}{\lambda}\right)-\mathrm{tanh}\left(\frac{x-230L_0}{\lambda}\right)\right]    , 
\end{equation}
\begin{equation}
b_{yb}=-0.8b_0-\frac{100L_0(x-x_0)b_1 f}{[(x-x_0)^2+(y-y_0)^2]}
 \left[\mathrm{tanh}\left(\frac{x-170L_0}{\lambda}\right)-\mathrm{tanh}\left(\frac{x-230L_0}{\lambda}\right)\right]    , 
\end{equation}
\noindent where $t\le t_{1}$ for $f=t/t_{1}$ and $t\ge t_{1}$ for $f=1$,$x_0=R_\mathrm{in}+200L_0$ and $y_0=H_0+6L_0$, $\lambda=0.5L_0$ , $ b_1=3.2\times10^{-4} $. Here $ t_1=300 $~s. The strength of the magnetic field below the bottom boundary will vary with time when $ t<t_1 $ and it will stop after $ t=t_1 $. The magnetic flux emergence by changing the conditions with time at the bottom boundary is set as the precious work\citep{2017ApJ...841...27N,2018RAA....18...45Z}.

\section{Numerical Result} % (fold)
\label{sec:Numerical Result}
Our numerical simulation work is mainly focused in the study of magnetic reconnection between the newly emerging magnetic field and the pre-existing magnetic field with different radial distances from the BH.  Here, we have studied the formation of the blob, their strength, and heating of the flow due to the magnetic reconnection for three cases as mentioned in the Table \ref{table1}. We find that the blobs/plasmoids formation occurred more frequently closer to the BH. The blob or fluid velocity and temperature is also increases with decreasing the radius of the simulation box. We also noticed that the matter velocity becomes super Alfv\'{e}nic along the current sheet and it also increases as matter moves up. These blob formation is very frequent in the simulation.  During the analysis in subsections \ref{sec:10}, \ref{sec:30}, \ref{sec:40}, the blob temperature is found mostly $>10^9$~K, which can be comparable to the observed temperature. The fluid velocity is achieved around ~$0.1c$ and it is also showing increasing trend with the fluid height.  Since our simulation length scale is very small, so there is a chance of other possibilities as well, for instance, the outflow velocity may decrease after a certain height or may form failed outflows. Moreover, since the magnetic reconnection process is heated the plasma above the disc surface, so it can form a hot corona. \newline

\begin{figure}
	\centering
	\includegraphics[width=0.80\textwidth, angle=0]{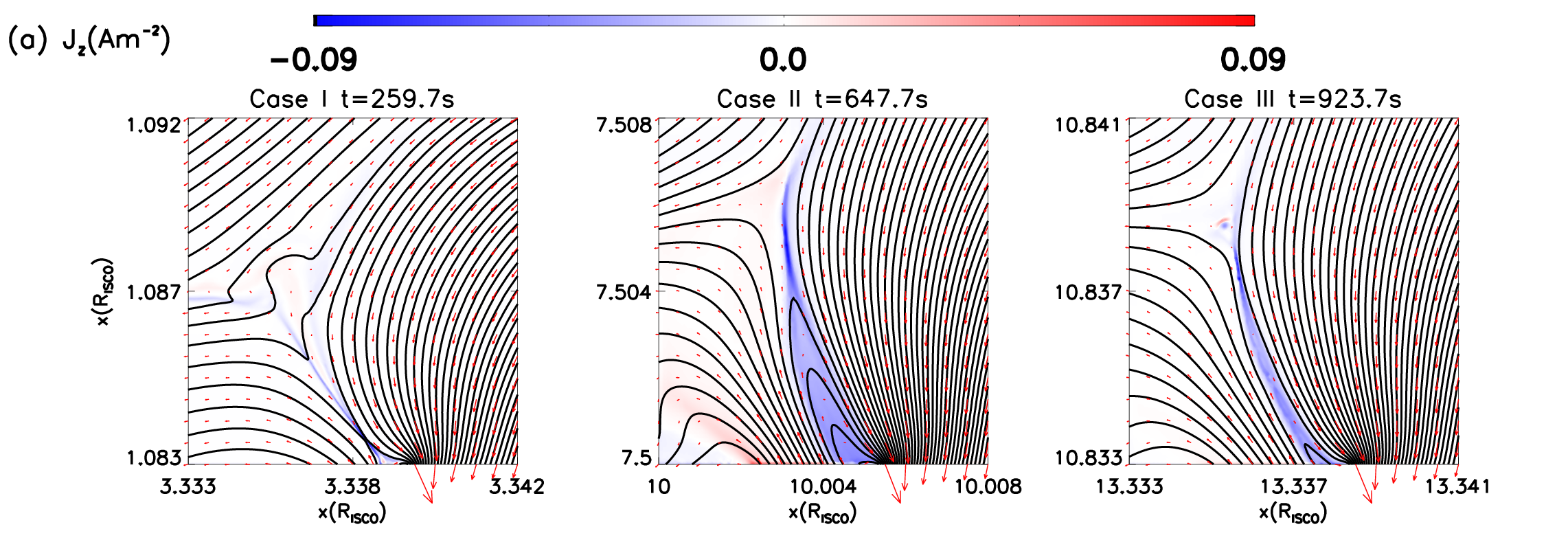}
    \includegraphics[width=0.68\textwidth, angle=0]{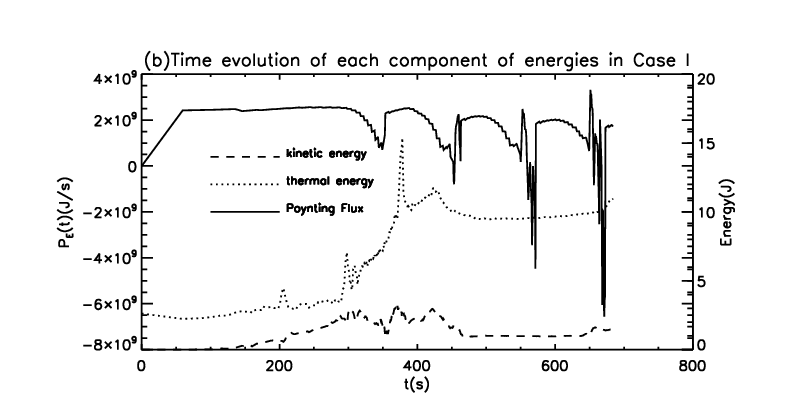}
    \includegraphics[width=0.68\textwidth, angle=0]{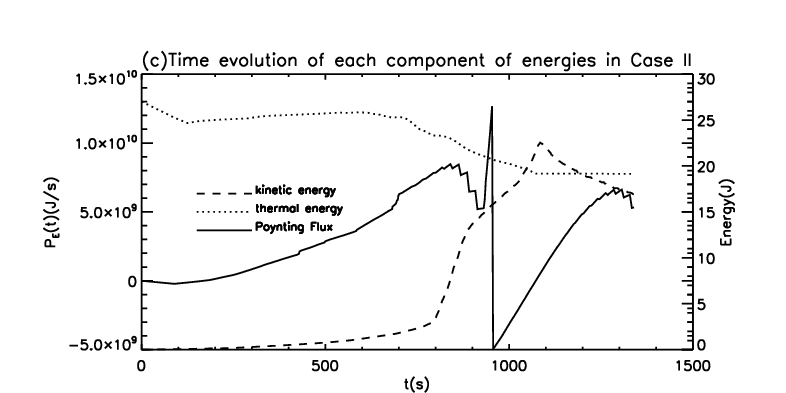}
    \includegraphics[width=0.68\textwidth, angle=0]{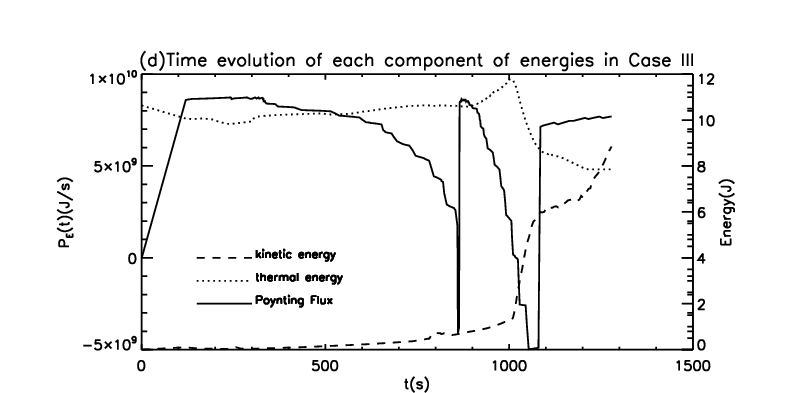}
	\caption{Comparison of the current density, $J_z$,  in three cases.  For Case~I $t=259$~s, Case~II $t=647$~s, Case~III $t=923$~s. The red arrows shows the direction of the magnetic field line. Time evolution of the magnetic power $P_\mathrm{E}$ in three cases is shown in (b)(c)(d), respectively.
	}
	\label{Figb}
\end{figure}

Figure \ref{Figb}(a) shows the current density distribution of the magnetic reconnection regions of all three cases.  The current density with positive and negative values represents the opposite directions of the motion. The current sheet is the main structure of the magnetic reconnection, so the current density can more intuitively reflect the shape of fluid. We plot the direction of the magnetic field line with the red arrows in most of
figures of the current density. From these arrows, the parallel and antiparallel lines are clearly distinguished. The fluid of Case I have turbulent
phenomena at the beginning, such as the magnetic island structure caused by the common tears-mode instability, the vortex-like plasma formed
by KHI, this phenomenon has lasted for a long time. In Case II, the turbulence is almost invisible, although it occurs at smaller scale. The fluid
is almost straight move to the top left-hand side than the current sheet of the Case I because of gravity effects and the structure is also thicker.
The fluid in Case~III has almost same structures as in the Case~II but the currect sheet structure is more straight to upward due to low gravity pull. %\deleted{is thinner at given time the turbulent is more obvious than in Case~II, but it is not as obvious as that in Case~I}. 
Figure~\ref{Figb}(b)-(d) show the temporal evolution of 
Poynting flux, kinetic energy, and thermal energy in three cases, respectively. 
%(the same as in \citet{2018RAA....18...45Z} section 3.2),
The value of the magnetic power $P_{E}$ is calculated as $P_{E}(t) = \int\!\!\! \int \textbf{P}(x,0,t) \cdot d\textbf{S}_{xz}$, where $\textbf{P}(x,0,t)$ is the Poynting flux vector through the bottom boundary and $d\textbf{S}_{xz}$ represents the area at the bottom boundary. The direction of $d\textbf{S}_{xz}$ is along the $y$-axis. Since all the variables are only functions of $x$, $y$ in space and they do not change along $z$-direction, we then assume $d\textbf{S}_{xz}=dxl_z\mathbf{\hat{e_y}}$ and $l_z=100L_0$\citep{2018RAA....18...45Z}.  
In subsections \ref{sec:10}, \ref{sec:30}, \ref{sec:40} we analyzed the shape and structure of each case in detail with the evolution of time, separately.\newline

\subsection{Case~I: Magnetic reconnection process at radius $10R_\mathrm{s}$ of the accretion disk }
\label{sec:10}
\begin{figure}
	\centering
	\includegraphics[width=\textwidth, angle=0]{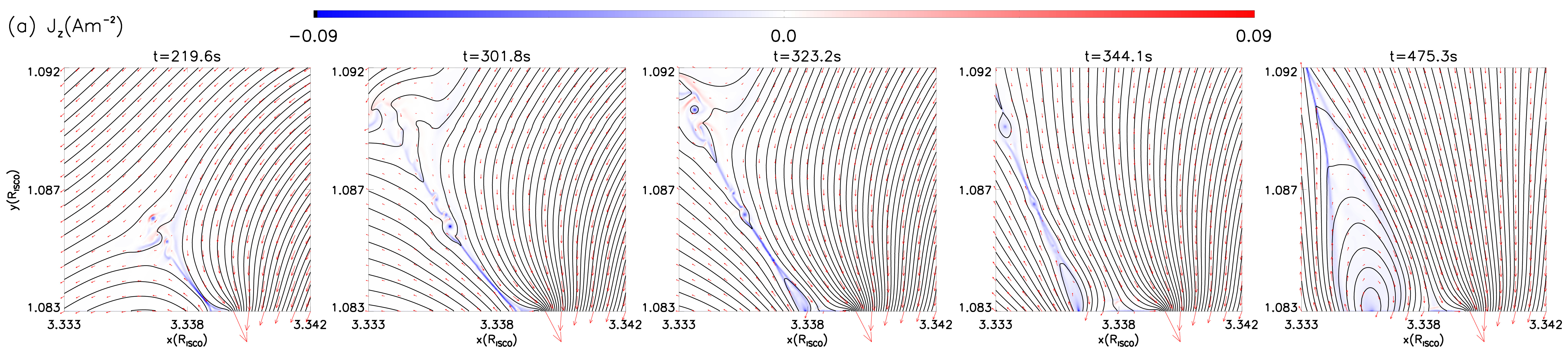}
	\includegraphics[width=\textwidth, angle=0]{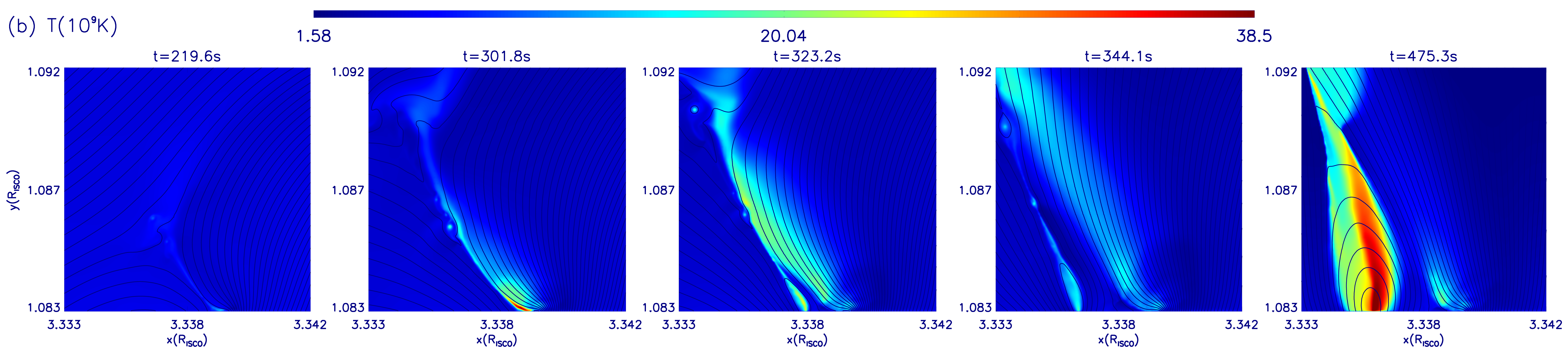}
	\includegraphics[width=\textwidth, angle=0]{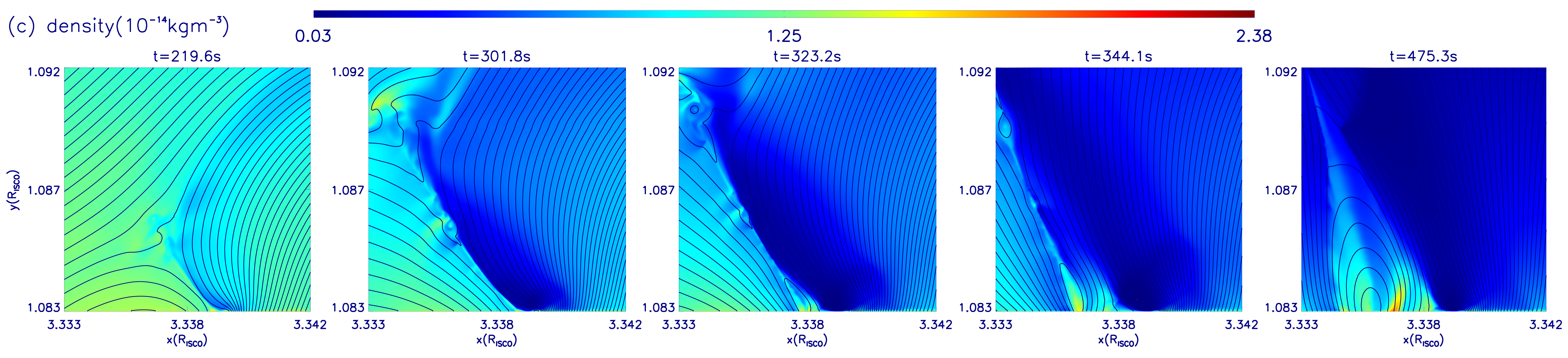}
	\includegraphics[width=\textwidth, angle=0]{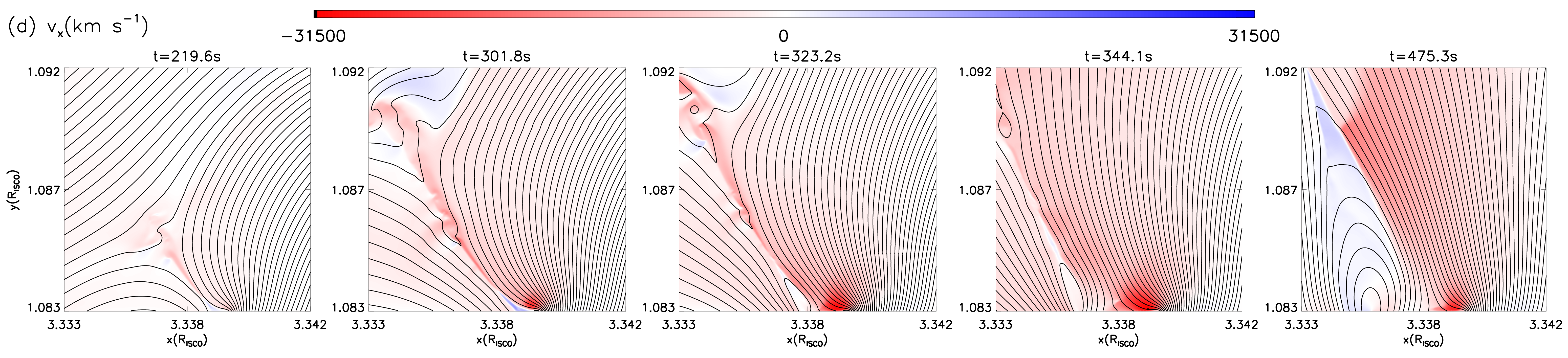}
	\includegraphics[width=\textwidth, angle=0]{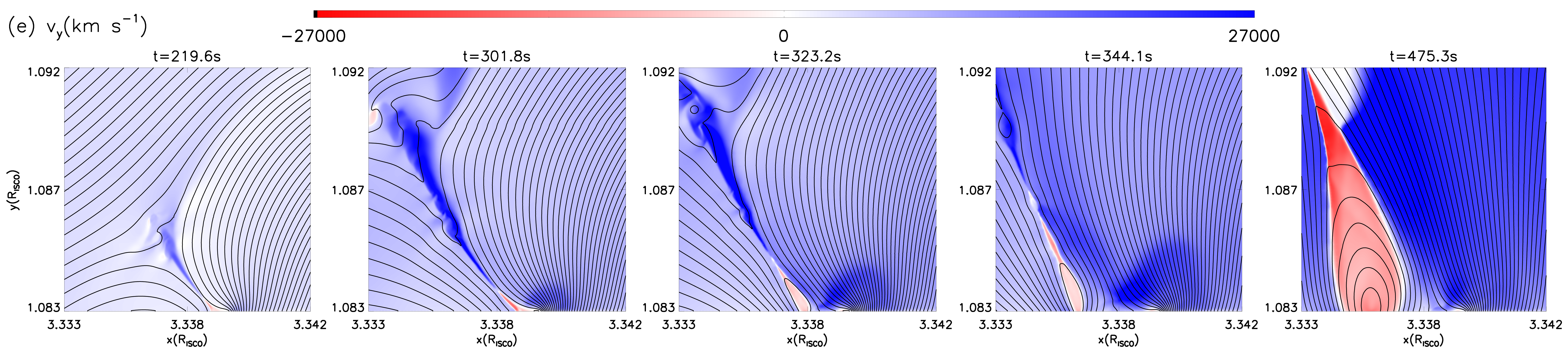}
	\caption{Distributions of different variables at five different times in Case I, (a) Current density, $J_z$, the red arrows show the direction of the magnetic field line, (b) Temperature, $ T $, (c) Mass density of plasma, $density$, (d) Velocity of the x-direction, $v_{x}$, (e) Velocity of the y-direction, $v_{y}$. Solid black lines represent the magnetic fields.}
	\label{Fig1}
\end{figure}

%\deleted{The magnetic reconnection is occurred at $10R_\mathrm{s}$ on accretion disk from the black hole in Case~I, and the density of disk equatorial surface is the biggest in these three Cases. However, the simulation region is located above the accretion disk and the density in this region is lower than other cases as calculated by Eq\ref{rho}.} 
The rows of Fig \ref{Fig1} represents the distribution of the current density, temperature, mass density and velocity components in the $x$-direction and $y$-direction of the fluid at five different times (mentioned at top middle of each box).  The values of flow variables are represented by the colour bar at top of each row. From the picture, we can see that at the beginning of the reconnection a current sheet, pointing towards the upper left-hand side, is formed by the reconnection between the emerging magnetic field lines from the disc and the pre-existing magnetic field lines (straight diagonal lines in the box). After reconnection, field lines changed the shape and may become circular, vortex, or semi-loop like structure, which may form blobs/plasmoids or islands in the flow. 
The current sheet is pulled towards the left as soon as the arch appears because of the gravitational force. In these several hundred seconds, the fluid has different velocities due to both gravitational and magnetic forces, which can produced KHI because of the shearing between the different velocities as shown in the rows (d) and (e). So in this period of time, a small blobs with vortices is formed in the upper left-hand corner (column second and third) by KHI and is ejected from the box with tear-mode instability (narrow current sheet behind the plasmoid). This vortex structure is also quickly disappears or moves out from the simulation area as seen in column fourth and fifth of the Fig \ref{Fig1}. In fourth column at $t= 344$~s, blobs/plasmoids like structures appeared in the current sheet with TMI. In the same column, the small magnetic loop structure also appeared at the bottom of the simulation. In next column (last), the blobs/plasmoids are moved out from the box and the magnetic loop structures are formed in the fluid with foot-point inside the disc. In this case, the plasma is accelerates to high speed in a short time with the rapid rise of temperature (the highest temperature is achieved $3.8\times10^{10}$~K).\newline

\begin{figure}
	\centering
	\includegraphics[width=0.8\textwidth, angle=0]{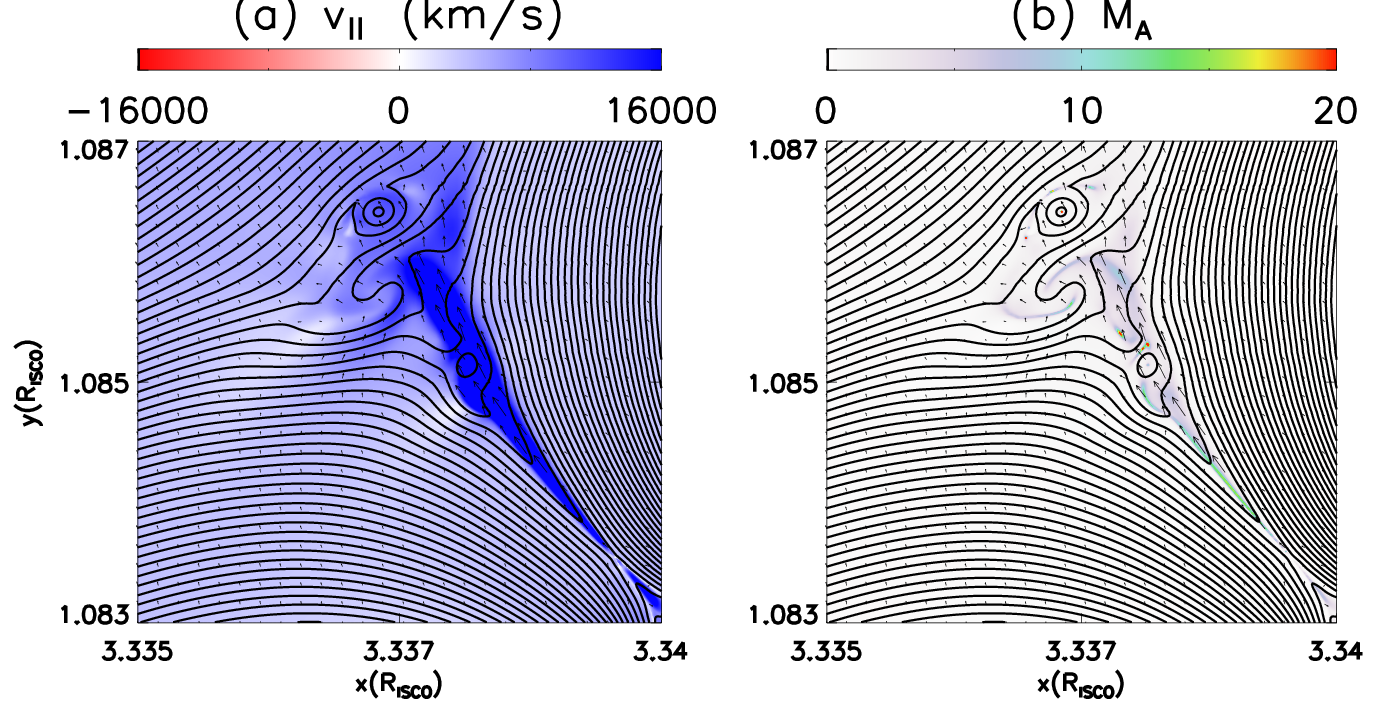}
	\includegraphics[width=0.8\textwidth, angle=0]{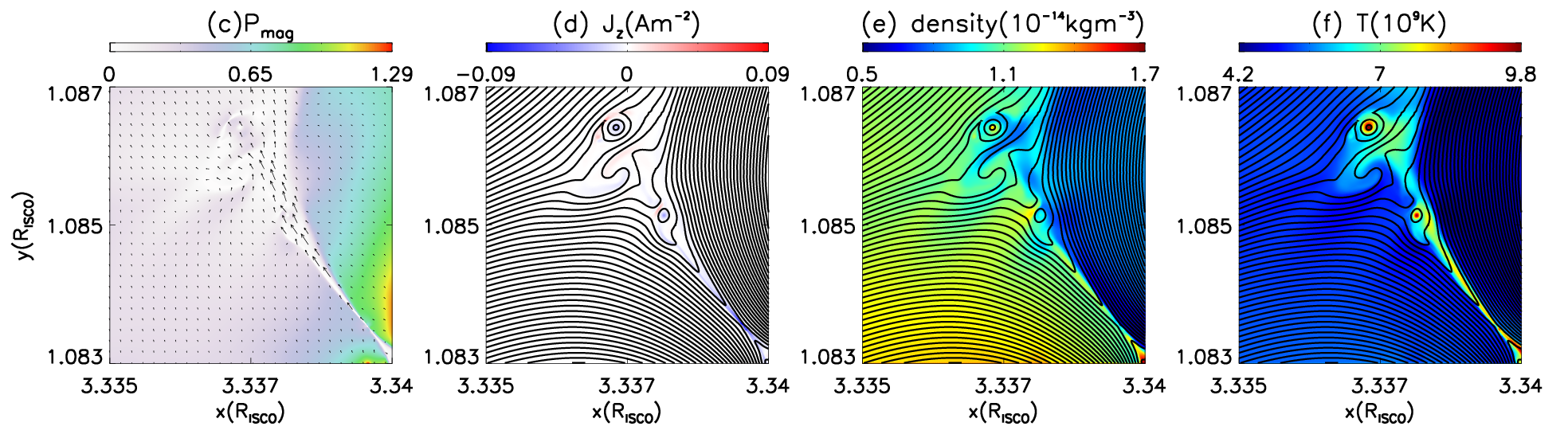}
	\caption{Distributions of  different variables of K-H instability at $t=217$~s. (a)Velocity of the fluid direction,$v_{||max}$, (b)Alfv\'{e}n mach number, $ M_A $, (c)Magnetic pressure, $P_b$, (d)Current density $J_z$, (e)mass density of plasma, $density$(f)Temperature, $ T $}
	\label{Fig2}
\end{figure}

\begin{figure}
	\centering
	\includegraphics[width=0.8\textwidth, angle=0]{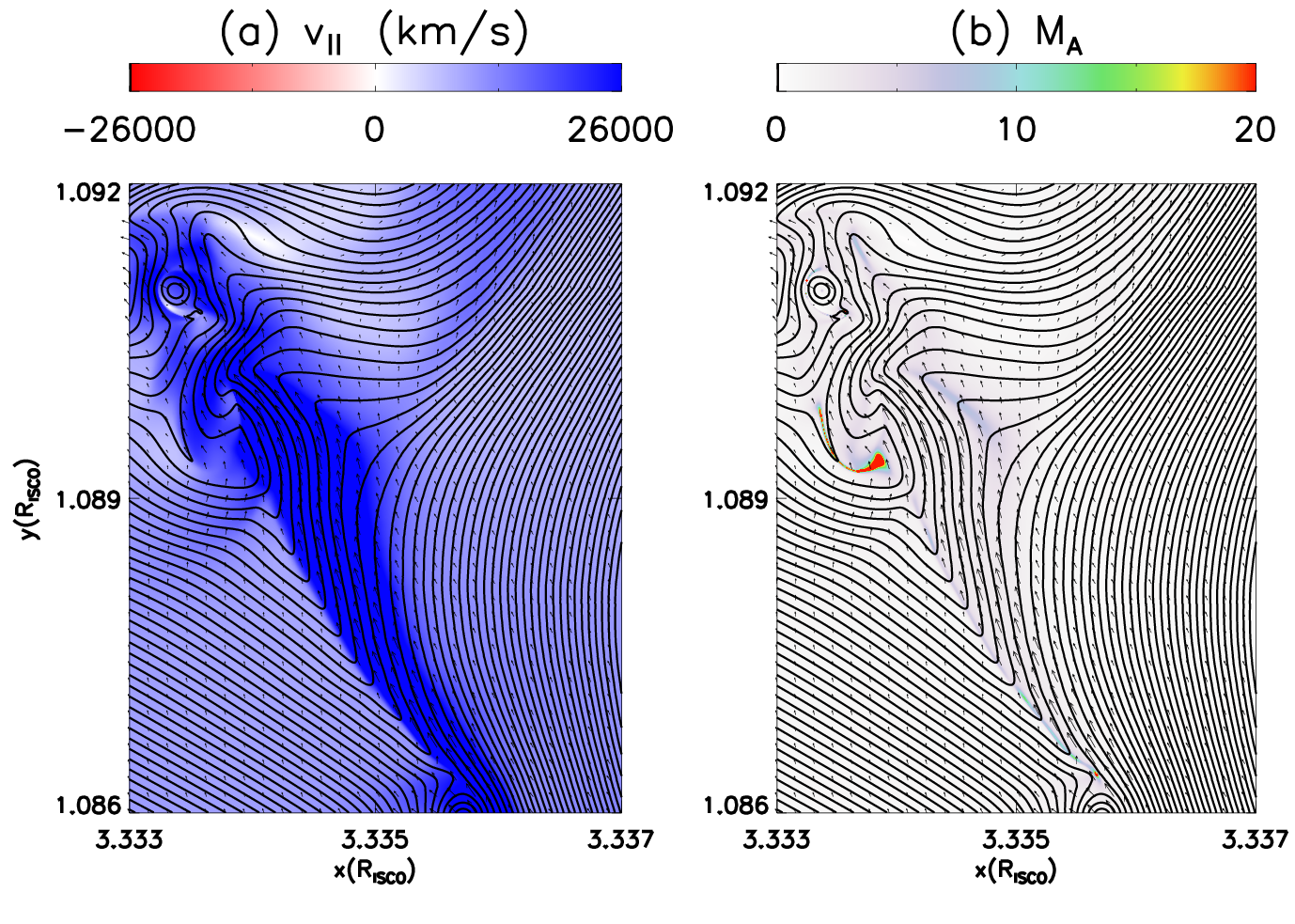}
	\includegraphics[width=0.8\textwidth, angle=0]{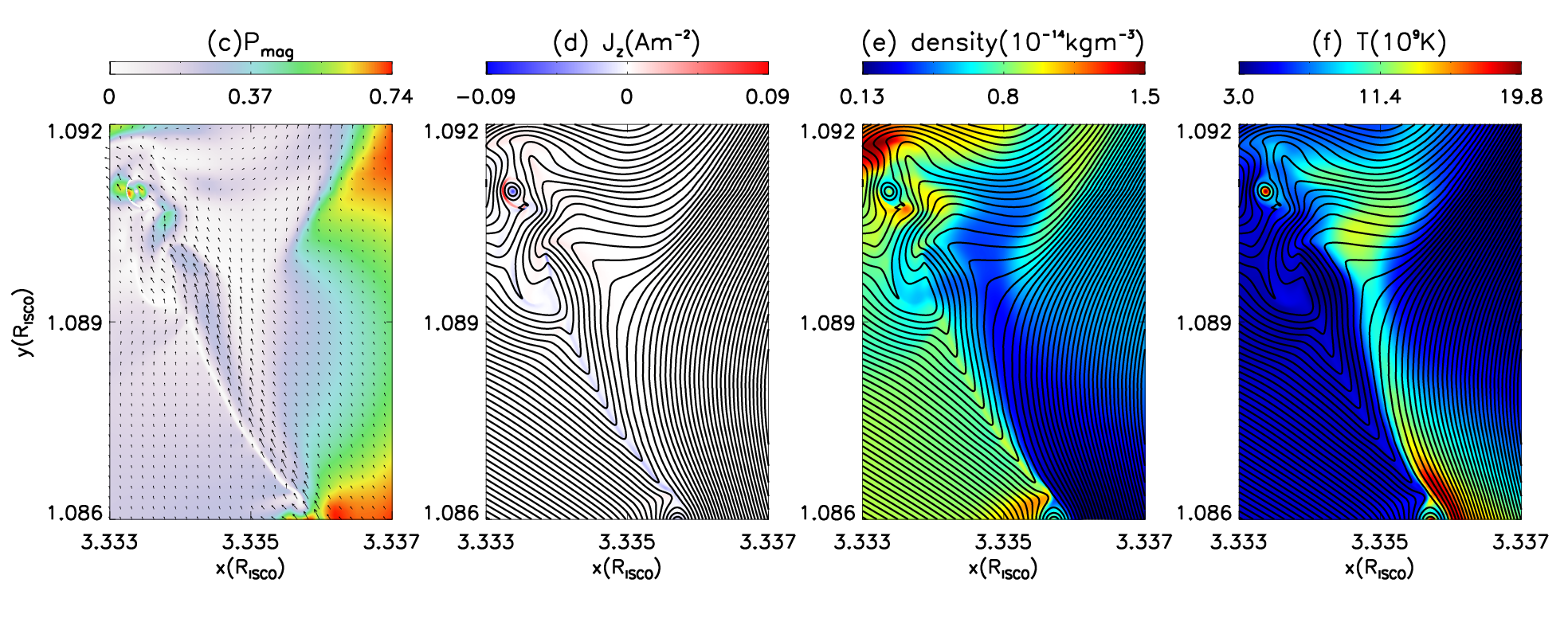}
	\caption{Distributions of  different variables of K-H instability at $t=314$~s.(a)Velocity of the fluid direction, $v_{||max}$, (b)Alfv\'{e}n mach number, $ M_A $, (c)Magnetic pressure, $P_b$, (d)Current density, $J_z$, (e)Mass density of the plasma, $density$ ,(f)Temperature, $ T $}
	\label{Fig3}
\end{figure}

\begin{figure}
	\centering
	\includegraphics[width=0.8\textwidth, angle=0]{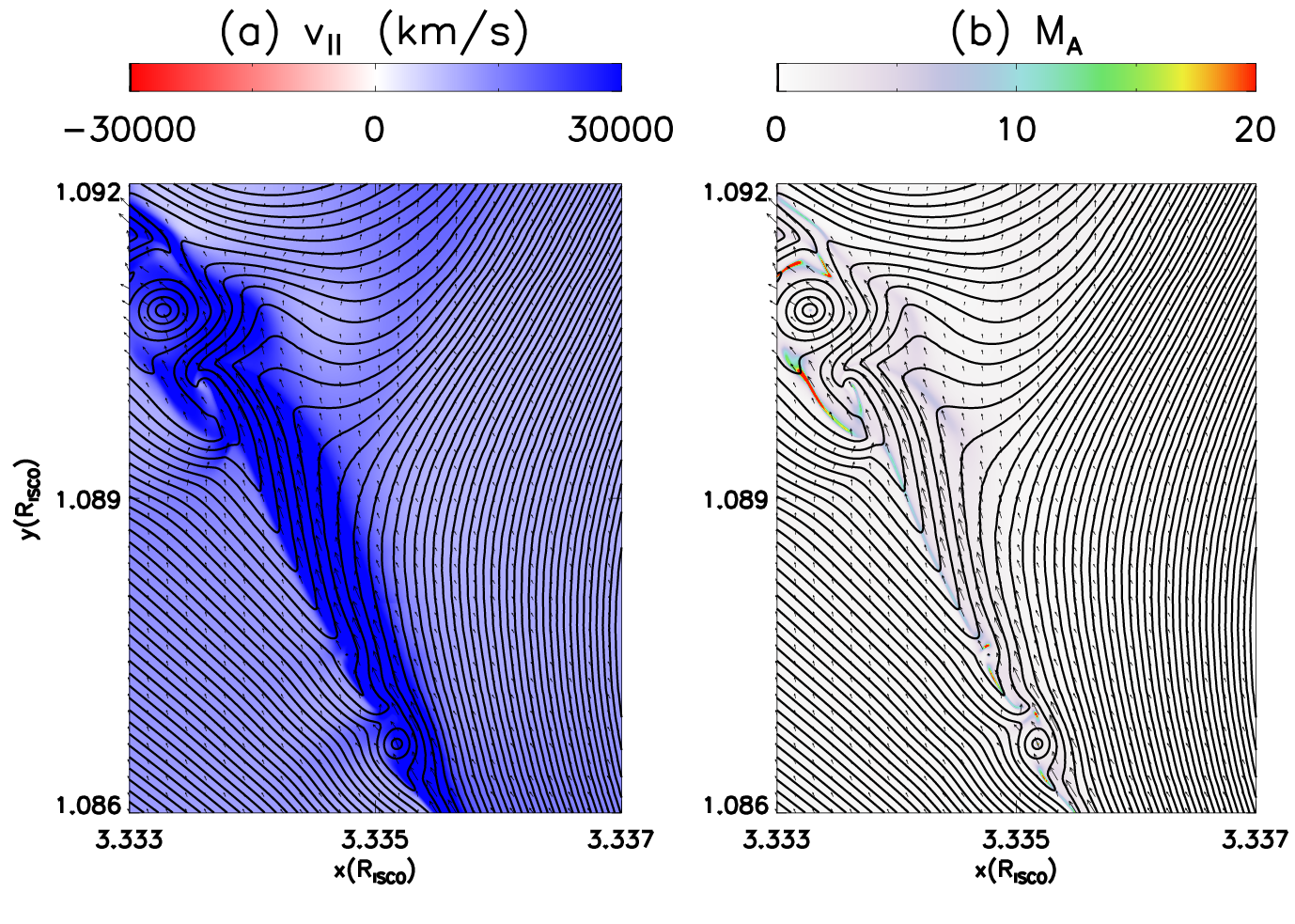}
	\includegraphics[width=0.8\textwidth, angle=0]{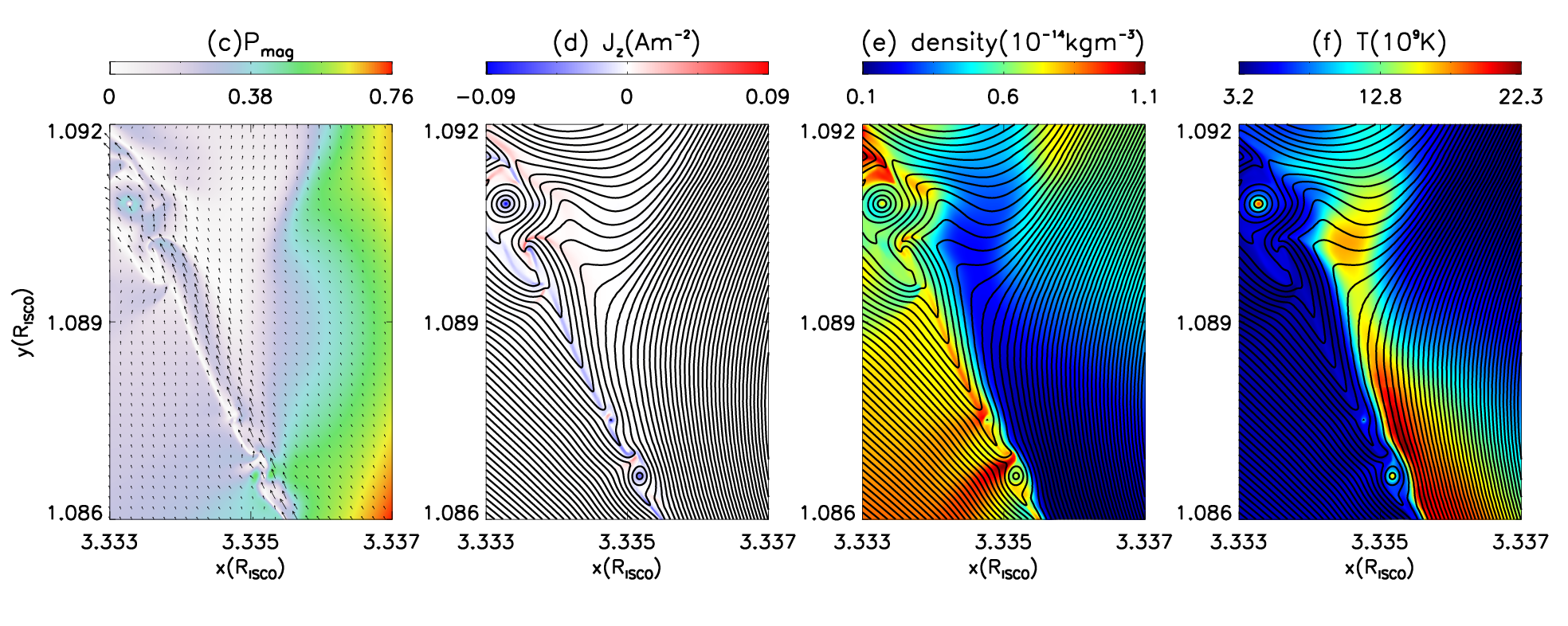}
	\caption{Distributions of  different variables of K-H instability at $t=324$~s. (a)Velocity of the fluid direction, $v_{||max}$, (b) Alfv\'{e}n  mach number, $ M_A $,(c)Magnetic pressure, $P_b$,(d)Current density, $J_z$, (e)Mass density of the plasma, $density$, (f)Temperature,$ T $}
	\label{Fig4}
\end{figure}

This is the nearest case to the black hole. Under the effect of the strongest gravitational field, the process of magnetic reconnection is fast comparing with other cases, 
so the efficiency of converting magnetic energy into heat energy and kinetic energy is also high. 
The highest plasma velocity is achieved nearly $ 0.1c $. In the process of reconnection, shear fluids with different velocities are formed. 
So the vortex-like blobs are formed by KHI and the magnetic island is formed by tears-mode instability can be seen almost all the time. The physical distribution of KHI on $t=217$~s, $t=314$~s, $t=324$~s, are showed in Fig \ref{Fig2}-\ref{Fig4}; 
here panels (a) and (b) shows the velocity and Alfv\'{e}n Mach number distribution with velocity vector field, respectively. 
The second row with panels (c), (d), (e), and (f) of Fig  \ref{Fig2}-\ref{Fig4} represents the distribution of magnetic pressure ($P_\mathrm{mag}$) with 
velocity vectors, current density, gas density and temperature, respectively. The evolution of K-H instability of these three periods is the most significant. 
In Fig \ref{Fig2} at $t=217$~s, the K-H instability is 
occurred and located around the middle right of the simulation box, at this time the height of plasma structure is lower and having the lower temperature and velocity. The foot-point of the arc is located at the right corner. In the panel (f), the two distinct high temperature blobs are displayed.
At $t=314$~s, Fig \ref{Fig3} is drawn with higher height and lower horizontal scale than the Fig \ref{Fig2}. 
In the Fig \ref{Fig3}, the fluid structure is evolved and moved to the upper left-hand corner of the box and reached to a relatively high position.  
The second blob like structure is appearing with complex semi-loop magnetic field structures at the bottom. 
%The foot-point of this structure may lie in the inflow disk region. 
%\deleted{the two distinct high temperature blob (there is a blob cut in half by 
%the drawing area directly below) and complex rolled up magnetic field structures are displayed in the Figure } 
At this time the highest temperature is 
achieved $1.98\times10^{10}$~K, and the highest speed is $ 26000 $~km~$s^{-1}$. 
At $t=324$~s, the fluid is continued to evolve with their complex magnetic field structures and moving upward. 
The temperature and the plasma velocity along $y$-direction is still rising with their corresponding highest values $2.23\times10^{10}$~K, 
and $ 30000 $~km~$s^{-1}$, respectively. 
In panels (c) of the Figs  \ref{Fig2}-\ref{Fig4}, the overall trend of $P_{mag}$ is decreases with the height in the simulation box, 
so this help to accelerate matter to move upward and $P_{mag}$ is lower along the current sheet. 
%Since the $P_{mag}$ is high in the right side of the main current sheet, therefore the plasma blobs and current sheet moves from right lower to left upper corner. 
Panels (d) of the Figs  \ref{Fig2}-\ref{Fig4}, the $J_z$ is increases with the time and mostly founds in the active regions. 
Panels (e) of same Figure , the gas density distribution is complex but it is mostly higher in the lower left corner of the boxes but overall it is decreasing with height. %, if we look at the color bar of the each nnn 
The Fig \ref{Fig4} has two blobs, and both will be ejected one by one, so it may form blob like structure as move up. 
%(Figure  (\ref{Fig2}-\ref{Fig4}) are having three blobs). 
The Alfv\'{e}n Mach number ($ M_A=v_{||}/V_A $) is represented in the panels (b), where $v_{||}^2=v_x^2+v_y^2$ is bulk velocity of plasma.
The values $M_A$ is about 5-20  during the K-H instabilities, and %are formed, 
some of the local regions has more than 20. As we mentioned in the previous article \citep{2018RAA....18...45Z}, the KHI can be generated under this condition.\newline

\begin{figure}
	\centering
	\includegraphics[width=0.9\textwidth, angle=0]{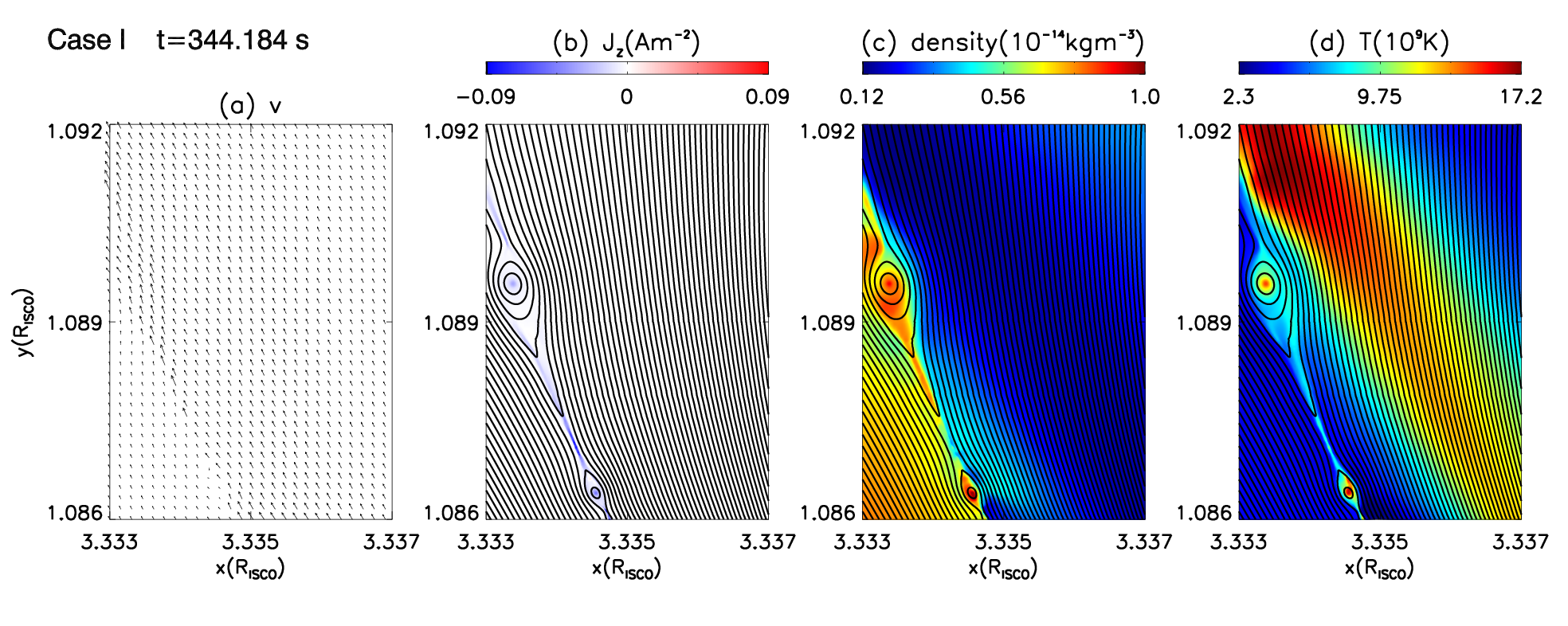}
	\caption{Distributions of  different variables of the big blob form at $t=344$~s.(a)Velocity vector,    $\mathbf{v}$, (b)Current density, $J_z$, (c)Mass density of the plasma, $density$, (d)Temperature, $ T $.}
	\label{Fig5}
\end{figure}

Fig \ref{Fig5} shows the distribution of the velocity vectors, the current density, 
the mass density of the plasma and the temperature with two blobs structure in the flow at $t=344.184$~s, which is the zoomed version of fourth column of the Fig \ref{Fig1}. Both blobs are separating out with the tears-mode instability. 
The blobs have higher temperature and mass density than the local surrounding environment. Interestingly, during this time slot, the magnetic field structure is not much complex as in the previous Figs  \ref{Fig2}-\ref{Fig4}. This happens may be due to one direction current density as represented in the panel (b) of the Fig \ref{Fig5}.\newline

\begin{figure}
	\centering
	\includegraphics[width=0.9\textwidth, angle=0]{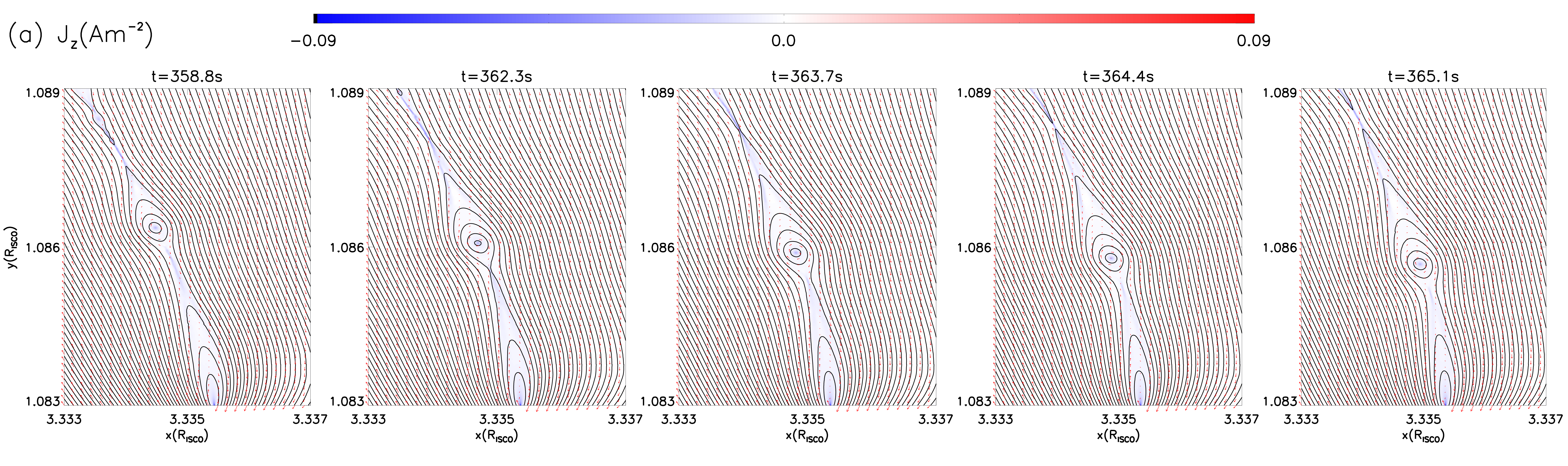}
	\includegraphics[width=0.9\textwidth, angle=0]{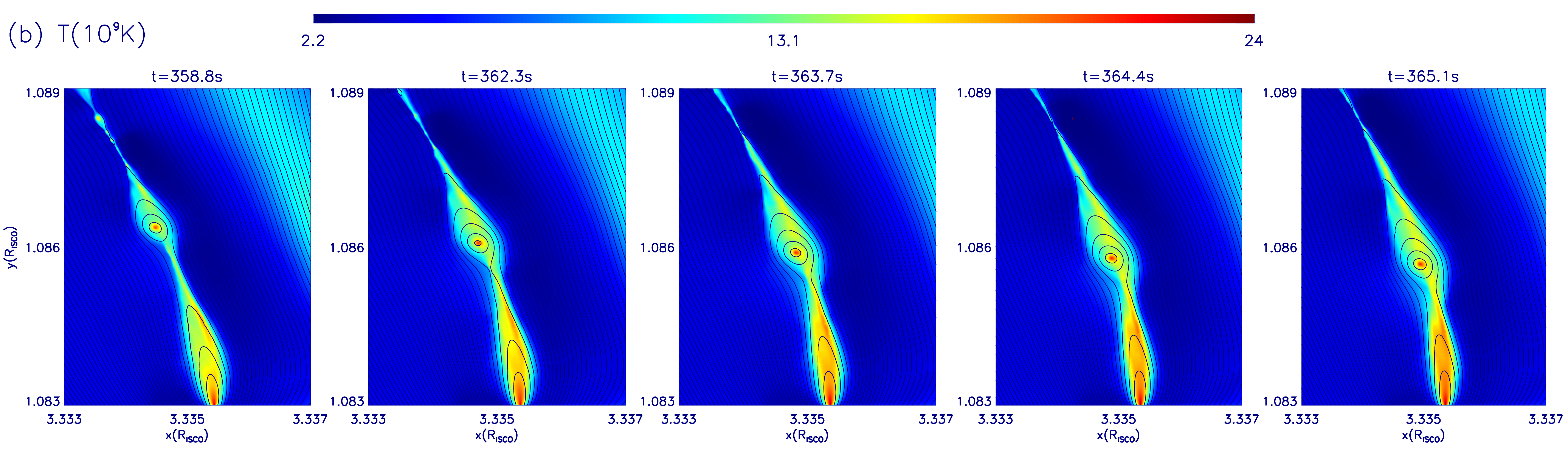}
	\caption{Distributions of different variables at five different times when the new reconnection process after blob is ejected (a) Current density, $J_z$, the red arrows is the vector direction of the magnetic field line, (b)Temperature, $ T $.}
	\label{Fig6}
\end{figure}

The upper larger blob appears (see Fig \ref{Fig5}) and moves out with tear-mode instability (upper left corner of Fig \ref{Fig6}) from the box, and other small lower blob come into the centre of the picture and their one circular magnetic line is reconnecting with newly emerging magnetic line from the disc as represented in the Fig \ref{Fig6}. In this time period, the arch like structure is formed having tear-mode instability at upper end and reconnetion process at other lower end, simultaneously. The current density and the temperature distribution of the new reconnection process and the new magnetic arch has been shown in the Fig \ref{Fig6}. %In Figure \ref{Fig6} 
The new magnetic arch has a higher temperature and unidirectional current density. The whole activities in Case I can be summarized as: some large blobs move out and the magnetic arch appears with the many consecutive reconnections, which formed a larger blob structure as presented in the last panel ($t=475.3s$) of the Fig \ref{Fig1}. 
A interesting thing is that there is initiated another side for reconnection beside the large magnetic loops structure at the bottom of last panel ($t=475.3s$) 
of the Fig \ref{Fig1}. It means the reconnection and blob formation are catastrophic processes, which is also mentioned in the theoretical model of \cite{2009MNRAS.395.2183Y}.  %dependent on the cascading time scale. 
Since the simulation box size is very small therefore many blobs/plasmoids can not come in one picture and they ejected out from the boxes. The even more larger blobs can be seen in the other two cases and related physical mechanism will also be discussed in the next subsection.\newline

\subsection{Case~II: Magnetic reconnection process at radius $30R_\mathrm{s}$ of the accretion disk }
\label{sec:30}
\begin{figure}
	\centering
	\includegraphics[width=\textwidth, angle=0]{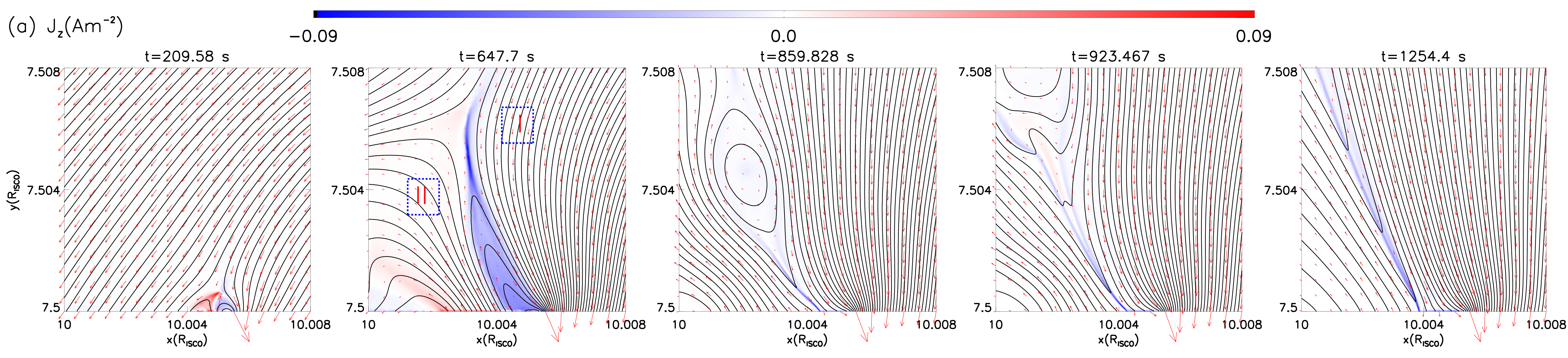}
	\includegraphics[width=\textwidth, angle=0]{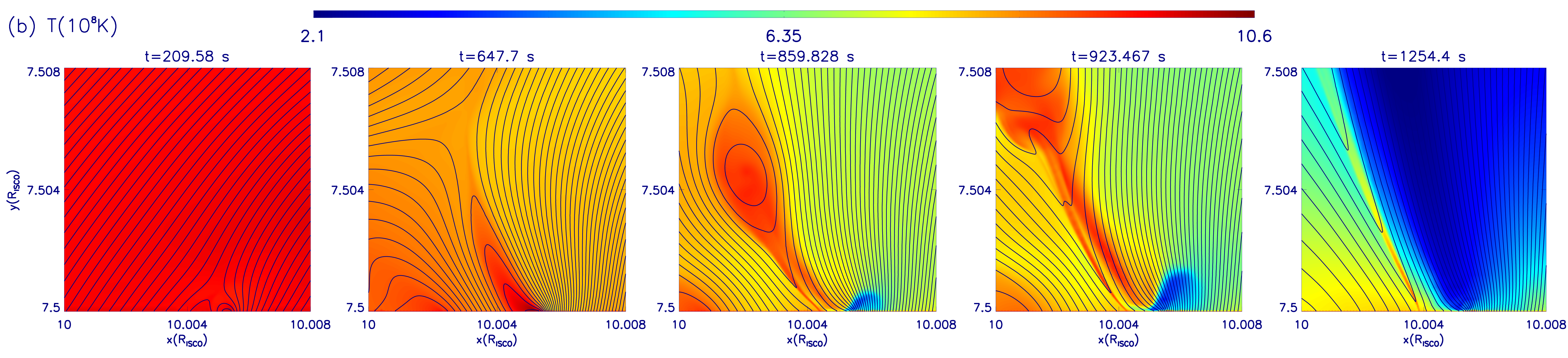}
	\includegraphics[width=\textwidth, angle=0]{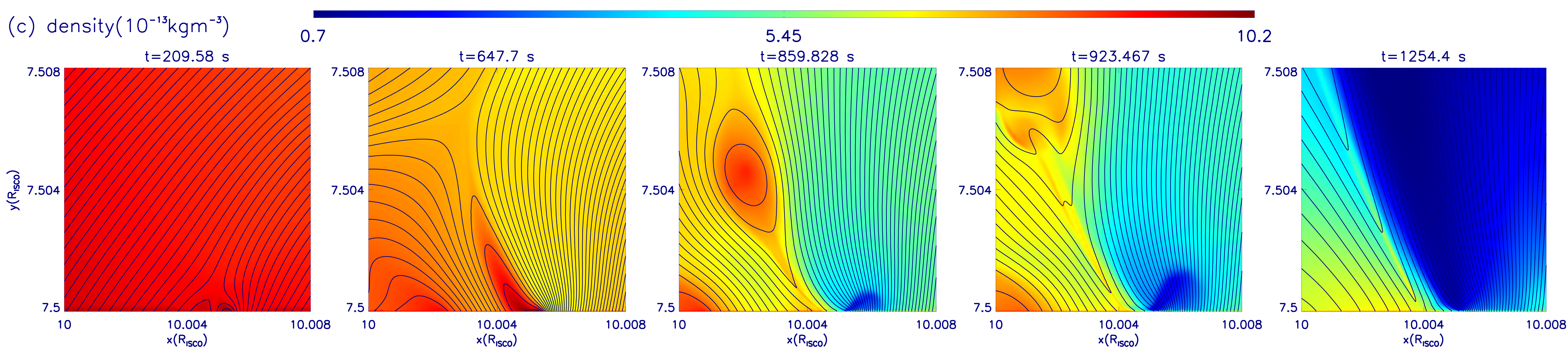}
	\includegraphics[width=\textwidth, angle=0]{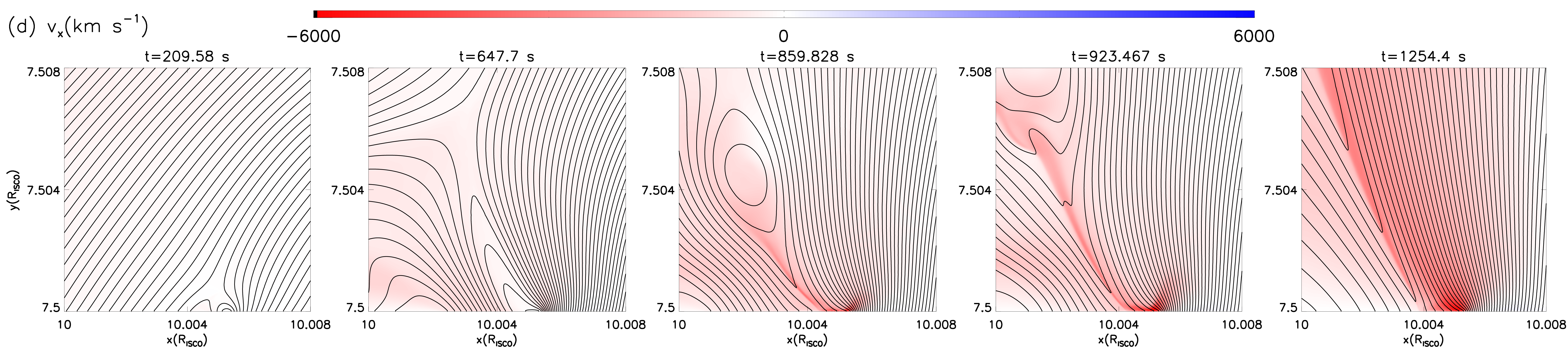}
	\includegraphics[width=\textwidth, angle=0]{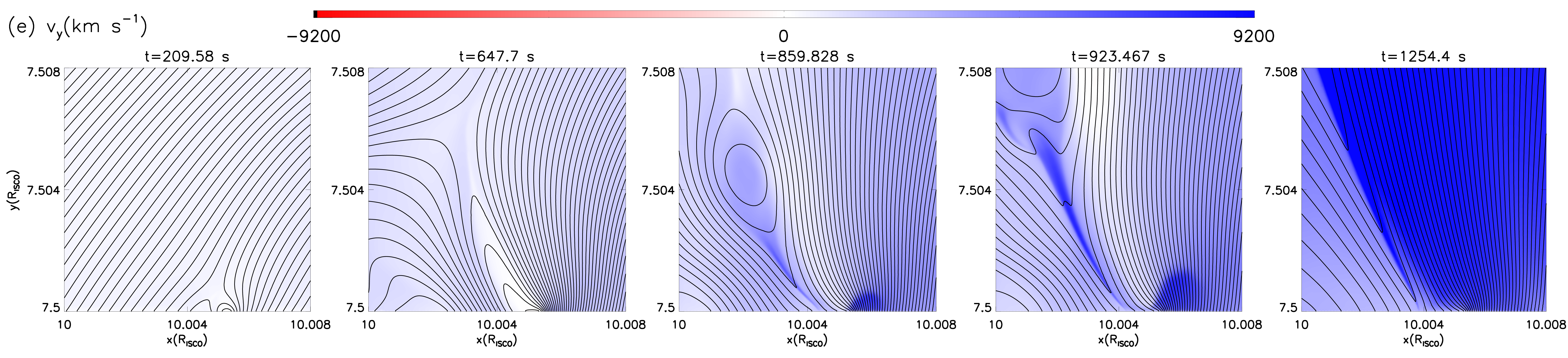}
	\caption{Same as Fig. 1 but for Case II.}
	\label{Fig7}
\end{figure}

Here we have simulated the MHD flow at the radius $30R_\mathrm{s}$ from the center of Sgr~A*'s accretion disk and we called it Case II. 
Fig \ref{Fig7} is displaying the distribution of the current density (row a), temperature (row b), mass density (row c), $v_x$ (row d) and $v_y$ (row e) 
with five different times. In this case, the blob is formed in bigger size than the Case I. This happen may be due comparatively low gravity and low flow velocity, 
which reduces the effect of tear-mode instability (stretching of flow lines) so field lines get time to merger and become bigger before move out from the bottom of 
the box. The horizontal movement of the current sheet is less and the current density is mostly unidirectional due to low turbulence in the flow. The field line structures are also not much complex than the previous case. So this blob structure is nicely looked as mentioned by \cite{2009MNRAS.395.2183Y}.
The maximum temperature of this bigger blob is reached around $1.06\times10^9$~K, which is lower than previous Case I. 
The plasma blob also have high density as is shown in the 
density distribution in Fig \ref{Fig7}. After the huge blob is eject at $t=923$~s. A thin current sheet is formed at the end of the fluid flow at $t=1254$~s, which still facilitate the reconnection process. \newline

\cite{2009MNRAS.395.2183Y} have explained the cause of intermittent jet. They analysed the reason for the corona mass ejection (CME) and concluded that the Lin–Forbes model of CME \citep{2000JGR...105.2375L} can be combined with the accretion disc of the BH. As It is described in their paper that the sheared accretion flow and turbulence can deform the magnetic field to form a flux ropes, then the energy can be accumulated and stored into magnetic field lines until the threshold value is reached; after that the system loses balance, the flux rope (blob) will be pushed out by the magnetic pressure, and a thin current sheet will appear below  (See columns 4 and 5 of the Fig \ref{Fig7}), which facilitates the reconnection process. In their theory model, the plasma can easily reach relativistic velocity with a large length scale and the jet is launched from very closed to the BH. Our simulation area is a lot smaller than  \cite{2009MNRAS.395.2183Y}, so the reconnection scale is smaller. The heat energy and the kinetic energy are converted by magnetic energy is also lower than \cite{2009MNRAS.395.2183Y}. The maximum velocity is not reached to the relativistic speed, since small simulation area also suppressed the evolution of blobs/plasmoids or disc structures with the distance. Although in this case, the maximum plasma speed is obtained $9200$~km~$s^{-1}$ close to the disc and it is more than the speed of the model on the solar corona \citep{2018RAA....18...45Z}. It is also larger than the speed of CME on the corona $2000$~km~$s^{-1}$ that referred by \citep{2009MNRAS.395.2183Y}. Therefore, it can be concluded that the efficiency of magnetic energy conversion into thermal and kinetic energy is also a lot higher than the common CME.\newline

\begin{figure}
	\centering
	\includegraphics[width=\textwidth, angle=0]{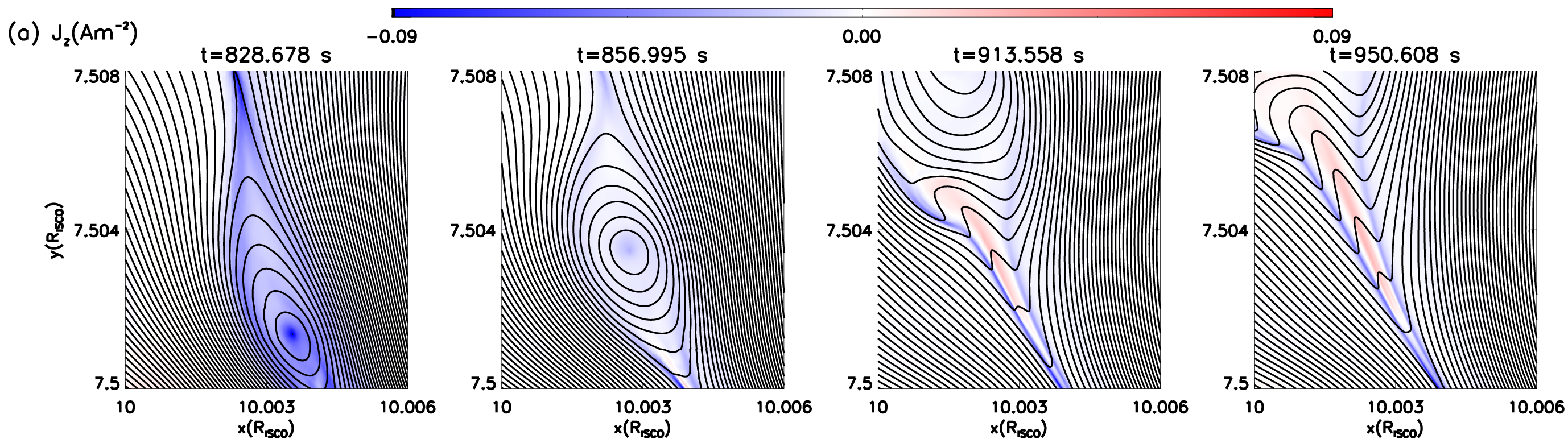}
	\includegraphics[width=\textwidth, angle=0]{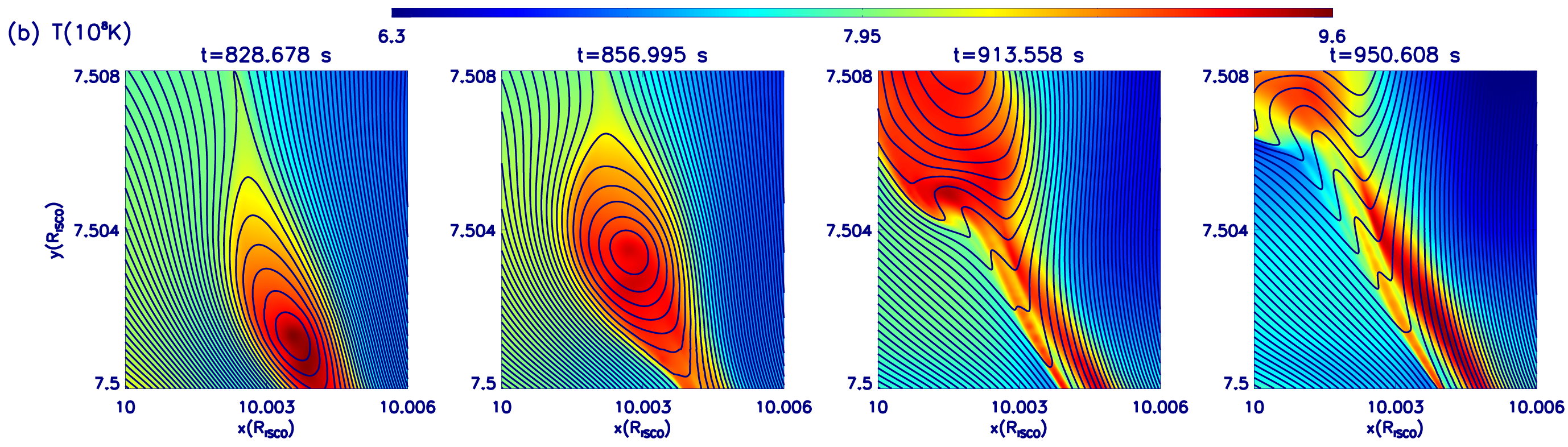}
	\includegraphics[width=\textwidth, angle=0]{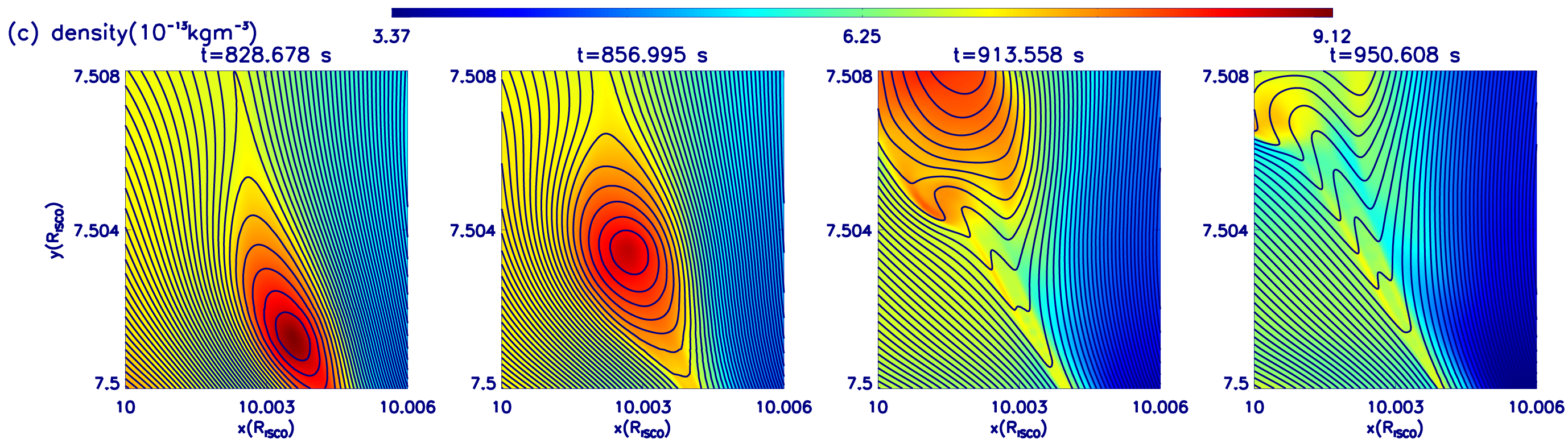}
	\caption{Distributions of different variables at four different times when the huge blobs appears in Case II.
		(a)Current density, $J_z$, (b) Temperature, $ T $, (c) Mass density of plasma, $density$. Continuous black curves represent the magnetic fields.}
	\label{Fig8}
\end{figure}

For the detail understanding of Lin-Forbes model and intermittent jet formation \citep{2009MNRAS.395.2183Y}, we have plotted Fig \ref{Fig8} with more  resolution of a particular event of the Fig \ref{Fig7}(the resolution level in Fig\ref{Fig7} is 3 and in Fig\ref{Fig8} is 5.). The Fig \ref{Fig8} shows the distributions of flow variables at four different times around the blob appears as in the Fig \ref{Fig7}. Here, we have more fine structures with higher resolution, which we could not see in the previous run for the Fig \ref{Fig7}. The blob has circular magnetic field loops, just like the projection of flux rope on two dimension. That is to say that the magnetic field structure of blob are satisfying the structure of flux rope in two dimension, as described by \cite{2009MNRAS.395.2183Y}, and the blob is caused by the turbulence of accretion flow and the tears-mode instability in the current sheet, which makes deformation in the magnetic field and forms a flux rope (the blob seen in the two-dimensional diagram), and it ejected by the magnetic pressure when the energy reaches the threshold value. In Case~II, the foot point of emerging flux is located below the surface of accretion disc. Because of the magnetic freezing effect, there will be convection turbulence to control the magnetic field during the process of magnetic flow emergence, which is also an important reason for the formation of the blob. Some scholars think that in the solar plasma environment, turbulent in the photo-sphere will lead to the formation of a flux rope (blob) in the corona, which can form solar prominence and filaments in the corona\citep{2009MNRAS.395.2183Y}. Therefore, BH accretion flow and the solar outflow components may have the similar characteristics \citep{2009MNRAS.395.2183Y}, which also shows that there is a common physical mechanism on the BH accretion disc. Even the accretion flow have the stronger turbulence. \newline

\begin{figure}
	\centering
	\includegraphics[width=0.4\textwidth, angle=0]{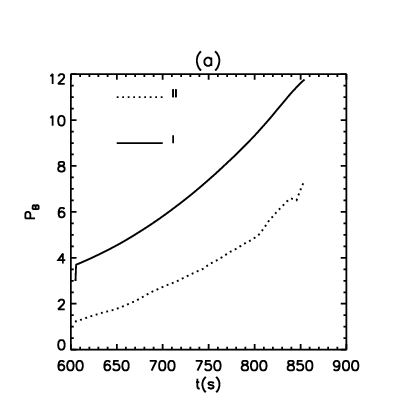}
	\includegraphics[width=0.4\textwidth, angle=0]{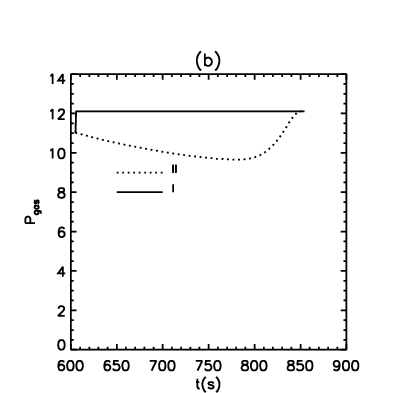}
	
	\caption{Distributions of different variables changes over time around the huge blobs appears in Case II.
		(a)Maximum magnetic pressure, $P_b$, (b) Average gas pressure, $P_{gas}$. }
	\label{Figc}
\end{figure}

In Fig \ref{Fig8}  the blob is formed by the reconnection from the magnetic field line on both sides of the current sheet. The magnetic reconnection is occurred when two opposite directions of magnetic lines come close to each other, break, and reconnect to form new structure and a magnetic neutral zone or neutral point is formed between them and it is called as X-point.  In Fig \ref{Fig8} there are two X-points formed at the both ends of the blob along the current sheet. The reconnection is formed between the surrounding magnetic line beside the current sheet due to pushing of the magnetic lines towards the current sheet from both side by originated plasma pressures. We calculate the magnetic pressure and gas pressure around the blob and location denoted as Area I and Area II (see second column of first row of the Fig \ref{Fig7}). corresponding to both locations, the magnetic pressure changes over time is shown in Fig.\ref{Figc}(a) and the gas pressure changes over time is shown in Fig.\ref{Figc}(b). Both the pressures in Area~I is higher than in Area~II and the magnetic pressure in both area are increasing in trend. So magnetic pressure and mass density gradient push magnetic line toward current sheet and increasing trend of magnetic pressure plays an important role in this scenario. Doing so, field lines come closer, then the reconnections occurred. Some people believed that there are countless number of small-scale magnetic reconnection happening in large-scale thick current sheet \citep{2009MNRAS.395.2183Y}. So flux rope breaks and moves outwards along the current sheet. Then, the magnetic field under the blob is stretched to form a thin current sheet. If the magnetic flux is threatened with the accreting flow then this reconnection process can occur in any region of the accretion disc. As we have seen that the reconnections happen faster close to the BH meaning the magnetic energy is converted faster into the kinetic energy and the thermal energy, so that the magnetic energy accumulated less in the plasma  \citep{2009MNRAS.395.2183Y}. So this is the reason the blob size in the Case I is smaller than the other two cases. \newline
	
	\subsection{Case~III: Magnetic reconnection process at radius $40R_\mathrm{s}$ of the accretion disk }
	\label{sec:40}
\begin{figure}
	\centering
	\includegraphics[width=\textwidth, angle=0]{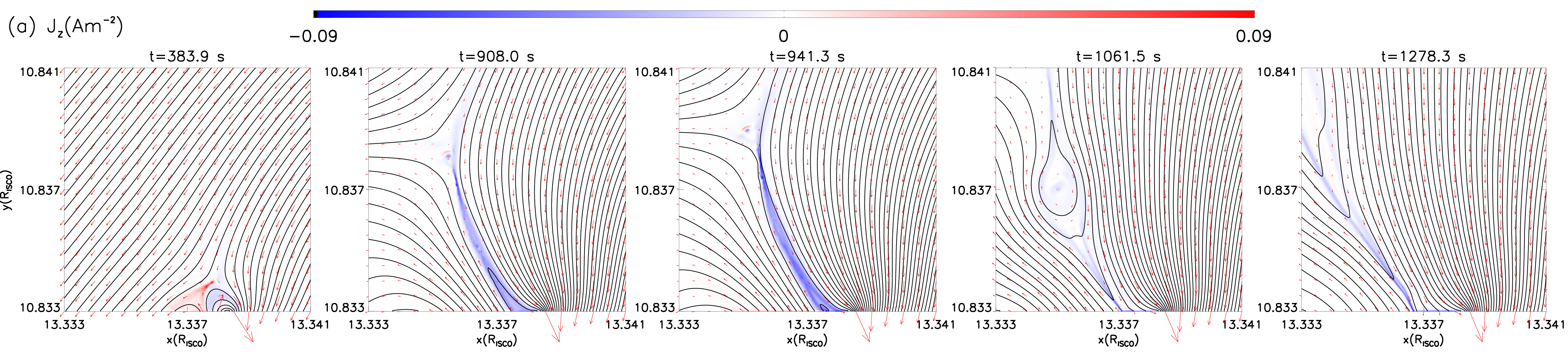}
	\includegraphics[width=\textwidth, angle=0]{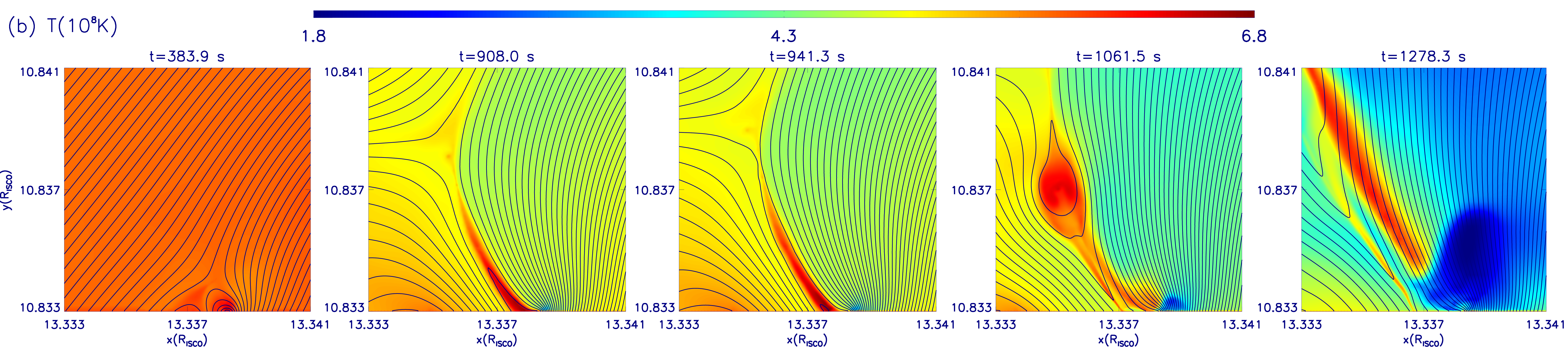}
	\includegraphics[width=\textwidth, angle=0]{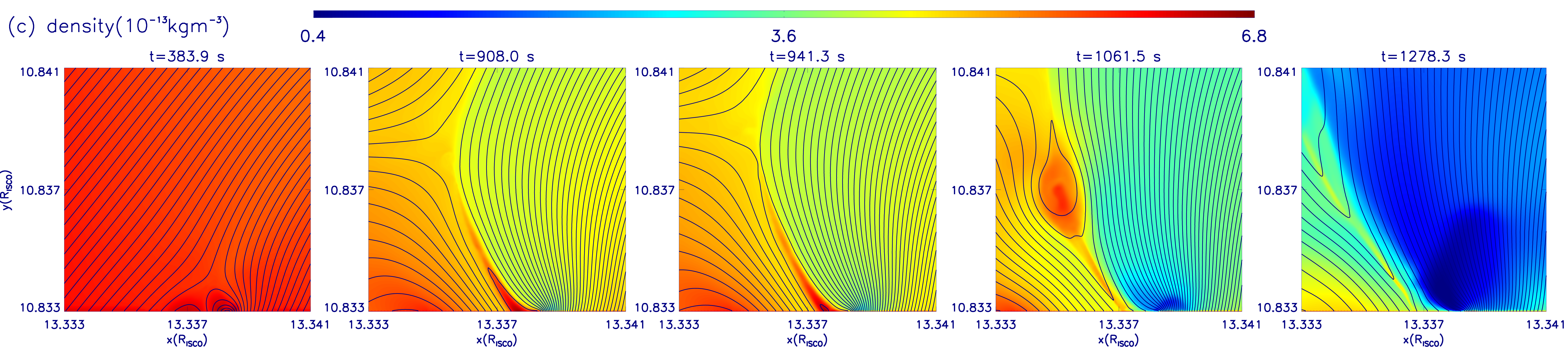}
	\includegraphics[width=\textwidth, angle=0]{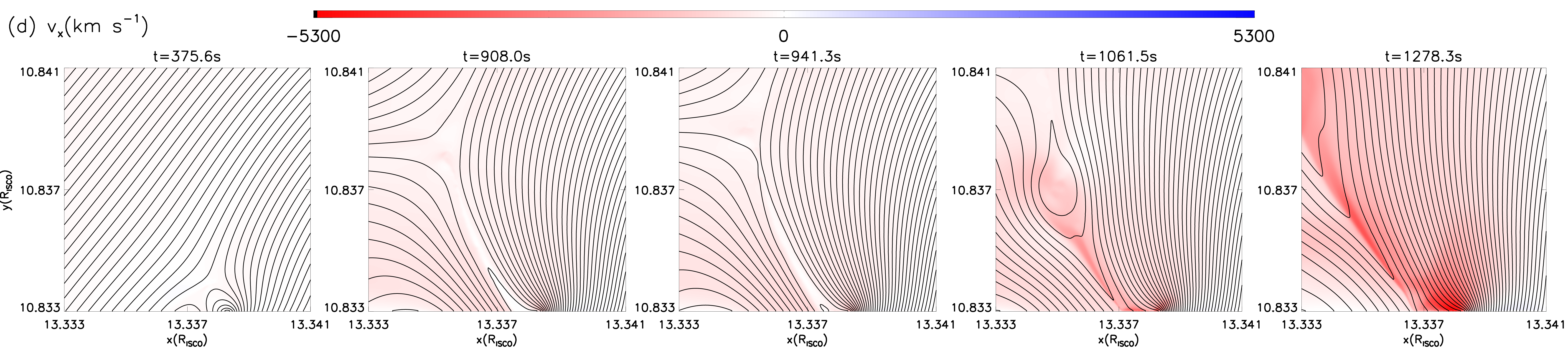}
	\includegraphics[width=\textwidth, angle=0]{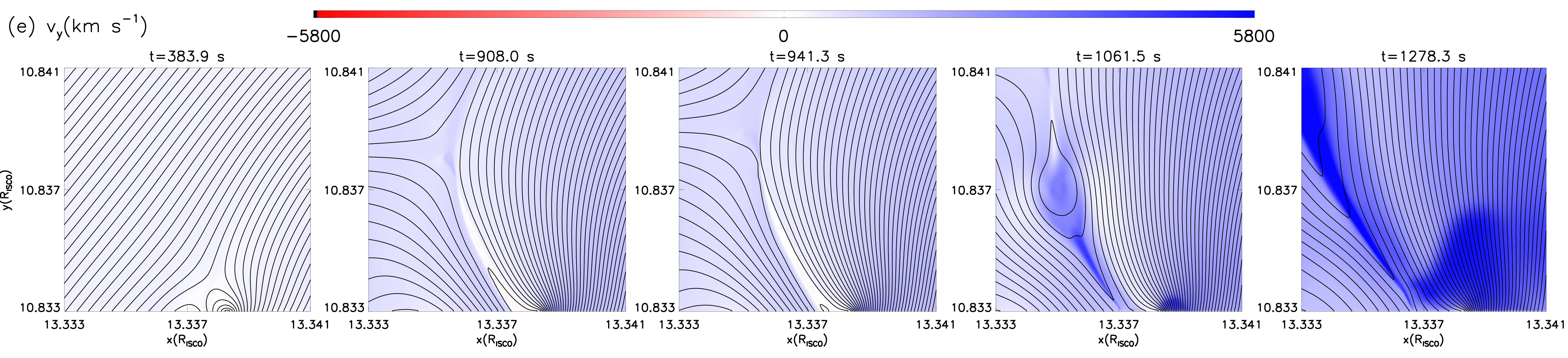}
	\caption{Same as Fig. 1 but for Case III.}
	\label{Fig9}
\end{figure}

	Case~III is the farthest from the BH of three cases, so the effect of the gravitational field is the lowest. The physical distribution of the current density, temperature, mass density, velocity in x direction and y direction with five different times is given in Fig \ref{Fig9}. At the beginning of the magnetic reconnection, the positive current sheet is tilted to the right-hand side and it lasted for a long time. This state lasts for nearly 400 s, which is longer than the other previous cases.
	At $t=941$~s, the main current sheet is still in a collimated state, but the vortices like blob with the small positive and negative current fragments is appeared in the current sheet. At $t=1061$~s, the big blob is formed with the reconnections and it appears later than the Case~II due to slow plasma activities with smaller gravitational field. This blob also collides with the front magnetic field lines and reconnects with them, so it grows in size. Finally, the whole blob is disappearing and making behind a narrow current sheet. The plasma velocity and temperature of Case~III is much lower than the other two cases because of slow reconnect rate. \newline
	
	Table~\ref{table1} shows that $g_x=-47087.009 $~$\mbox{m\,s}^{-2}$ in Case~I, $g_x=-4535.1341 $~$\mbox{m\,s}^{-2}$ in Case~II,  $g_x=-2507.5922 $~$\mbox{m\,s}^{-2}$ in Case~III. Here, we are going to briefly summarize all the cases. As in Case I, the fluid is quickly turned and attracted by the strong gravitational field and the KHI phenomenon is also produced by the strong shearing between the different fluid velocities. Since the current sheet is formed quickly in the Case I than in others, which helps to make many plasma activities faster, therefore, the formation of the blobs is faster and has high speed to move outward. This also favours the many observations which are showing that the blob formation occurs around a few to few tens of the Schwarzschild radii \citep{2020ApJ...891L..36G,2012sci...338..355D}.   \newline

\section{Conclusions} % (fold)
\label{sec:conclusions}
Based on the magnetic reconnection simulations by \citet{2017ApJ...841...27N} and \citet{2018RAA....18...45Z}, we embedded the plasma environment of the AGN disk. In this work we do detail analysis of the shape and the physical mechanism of the small scale reconnection at the different radii. We also study the effects of the gravitational field on the magnetic reconnection and blob/plasmoid formation. The main conclusions from our numerical simulations are as follows:\newline

1. Our model is based on the connection of the newly emerging magnetic field and the pre-existing magnetic field. The current sheet is
formed by the tear-mode instability (stretching), and therefore the reconnection, which converts the magnetic energy into thermal energy and
mechanical energy. Since the gravity is along the $x-$direction in the simulation box, therefore the current sheet is also stretched towards the top
left-hand side in the box. If we look at the Case I, which has more gravitional acceleration therefore it has more turbulent and sheared flow,
which makes the magnetic field also complex therefore the current density has more positive and negative values, as seen the Fig \ref{Fig1} than the
other two cases. Since the flow is turbulent, so there is more chance of formation of flux tubes and therefore the occurrence of the magnetic
reconnections. Thus, the flow of the Case~I is more hot and have higher velocities than others. We find that the formation of the blobs and
reconnection rate is higher close to the BH, which also supports the recent study has done by \cite{2020arXiv200304330R} and \citep{2020MNRAS.495.1549N}. %Ripperda et al. 2020 (arXiv:2003.04330).   
In our small scale simulation, we got maximum blob velocity along the current sheet around $0.1c$. Since the fluid velocity is having radial stratification as seen in observations \citep{2019apj...887..147P,2019ApJ...879...8S}, means it increases with increasing radial distance from the center. The time-scale of the  three real flare events mentioned in  \citet{2020arXiv200514251B} is about $30-60$~min, and the result of their numerical model based on these observed events show the plasmoid size is $L_\mathrm{spot}=\frac{1GM}{c^2} \approx 5.86 \times 10^9$~m and the velocity is around $v=0.5c$.
 In contrast, in our simulations, the blobs in Case~I is small, the size of the huge blob in Case~II and Case~III  is about  $5 \times 10^7 $~m. 
 However, combine with the Lin-Forbes\citep{2000JGR...105.2375L} model, 
 the newly formed magnetic arch with high temperature in the last column of the Fig. \ref{Fig1} in Case~I may also correspond to the observed flare events, and the magnetic arch size is about $5 \times 10^7 $~m. The maximum velocity in Case~I can reach $0.1c$, and the velocity in Case~II and Case~III are lower. The life time of the blobs in three cases is around $10-23$~min, 
 which is shorter than that in \citet{2020arXiv200514251B}. Therefore, the time-scale in Case~I is relatively close to the observation, but not those in Case~II and Case~III. Right now, our simulation box is very small, so we can not track the flow at very large distances, we will leave that for the our future study.\newline

2. In our simulation, we used the idea of the reconnection catastrophe model for blob formation that referred by \cite{2009MNRAS.395.2183Y}.  In all three cases, we have found that the magnetic reconnection process and blob/plasmoid formation in the disc corona are naturally occurring regardless of their gravity strength. But the reconnection and blob formation rates and their shapes and sizes are dependent on the KHI and tear-mode instability, which again depend on gravity and turbulence in the flow. Here, we got fairly high blob velocities than the CME velocities and the flow temperature is also very high, which is tally with the bright spots and episodic $X-$rays observations of the Sgr~$A^*$. So our simulation is supporting the theoretical model of blob eject formation can be possible by the magnetic reconnection process in the corona of the accretion disc \citep{2009MNRAS.395.2183Y}. %calculates by \citep{2009MNRAS.395.2183Y} can occur on the disk.
\newline

3. The magnetic reconnection processes close to the BH is faster, means the magnetic energy is more quickly converted to the kinetic and thermal energies. Therefore, the plasma speed and the temperature is higher in the Case I than the other two cases. But the blobs are smaller in size with vortices like shape and more frequently forms in the vicinity of the BH as seen in the Case I. This may happen due to strong KHI, which makes vortices like shape and strong tear-mode instability, which enhance frequent or episodic blob formation. Both the instabilities are strongly affected by the turbulent flows and gravity pull or stretching. Thus, the over all effect is that the magnetic energy accumulation is less in the magnetic flux tubes due to high reconnection rate and resulting the blobs are formed with smaller size than the low gravity pull cases.  
\newline

In this work, the magnetic reconnection model is embedded on the BH disc with using various input parameters of the disc flow then the associated behaviours of the small scale reconnection system in 2D have been investigated. We have done the study of small scale plasma activities assuming a constant gravitational acceleration in the small simulation box and checking the magnetic reconnection scenario for the intermittent jet formation as theoretical described by the \cite{2009MNRAS.395.2183Y}. In future, we have plan to do same study with large scale simulation box, means gravity will vary within the scales of the box. Doing so, we will hope that we can investigate the large scale relativistic jets with this reconnection scenario and other associated phenomena. This time, we have not considered the influence of electrons in the fluid but we are also hoping to consider it in future. Last but not the least, we will check this model with some more different initial boundary conditions, like input temperature, density, magnetic field structures, variable mass accretion rates, and so on. We also hope we can perform the true 3D numerical experiments in the future for looking into the formation of the flux rope (the counterpart of the magnetic island in 2D) during the magnetic reconnection.\newline

\section*{Acknowledgements}
We would like to thank the referee for the critical comments and suggestions, 
and Prof. Yanrong Li and Prof. Feng Yuan and Prof. Lei Ni for their helpful comments. 
This work is supported by National Natural Science Foundation of China (Grant No. 11725312, 11421303).
The numerical calculations in this paper have been done on supercomputing system in the Super-computing Center of University of Science and Technology of China. The figures in this paper are plotted on Super-computing in Prof. Wang,  Jun Xian's group.

\section*{Data availability}
The data underlying this article will be shared on reasonable request to the corresponding author.
%%%%%%%%%%%%%%%%%%%%%%%%%%%%%%%%%%%%%%%%%%%%%%%%%%
\part{title}


%\bibliographystyle{mnras}

\bibliographystyle{mnras}
\bibliography{blob}

\newpage
%%%%%%%%%%%%%%%%%%%%%%%%%%%%%%%%%%%%%%%%%%%%%%%%%%

%%%%%%%%%%%%%%%%% APPENDICES %%%%%%%%%%%%%%%%%%%%%

%%%%%%%%%%%%%%%%%%%%%%%%%%%%%%%%%%%%%%%%%%%%%%%%%%
\begin{table}
	\begin{center}
		\caption[1]{The parameters of the numerical  models.}
		\begin{tabular}{clclclclclclcl}
			\hline\noalign{\smallskip}
			Case  &  $g_x$~$(\mbox{m\,s}^{-2})$&$R_\mathrm{in}$~($R_\mathrm{s}$)& $\rho_0$~$(kg m^{-3})$ &$T_0(K)$&$H_0$&  plasma-$\beta$&$\sigma$               \\
			\hline\noalign{\smallskip}
			CaseI &-47087.009&
			$10R_\mathrm{s}$&$ 6\times10^{-10}$&$6\times10^{9}$&$3.25R_\mathrm{s}$&1.47596&0.00153402 \\ 
			CaseII &-4535.1341&
			$30R_\mathrm{s}$&$6\times10^{-12}$&$1\times10^{9}$&$22.5R_\mathrm{s}$&15.6385&$2.23868\times10^{-5}$\\
			CaseIII &-2507.5922&
			$40R_\mathrm{s}$&$3\times10^{-12}$&$0.6\times10^{9}$&$32.5R_\mathrm{s}$&6.08251&$3.43833\times10^{-5}$\\
			\hline\noalign{\smallskip}
			% new variable
			\label{table1}
		\end{tabular}
	\end{center}
\end{table}

	% typesetting comment
\label{lastpage}
\end{document}